\documentclass[referee,useAMS,usenatbib]{mn2e}
\def\lsim{\lower.5ex\hbox{$\; \buildrel < \over \sim \;$}}
\def\gsim{\lower.5ex\hbox{$\; \buildrel > \over \sim \;$}}
\def\be{\begin{equation}}
\def\ee{\end{equation}}

\def\bc{\begin{center}}
\def\ec{\end{center}}
\def\eg{{\it e.g.,}}
\def\etal{{\em et al.}}
\def\ie{{\em i.e.,}}
\def\ep{{e^--p^+}}
\def\el{{e^--e^+}}
\def\s{{\rm ps}}
\def\sk{{\rm sk}}
\def\kd{{\rm kd}}
\include{netbib.sty}

\usepackage{graphicx}
\title[Radiatively and thermally driven jets]
{Radiatively driven relativistic jets with variable adiabatic index
 equation of state}
\author[Vyas \etal]
{Mukesh K. Vyas$^{1}$, Rajiv Kumar$^{1}$, Samir Mandal$^{2}$, 
Indranil Chattopadhyay$^{1}$\\
$^{1}$Aryabhatta Research Institute of Observational Sciences 
(ARIES), Manora Peak, Nainital-263002, India\\
$^{2}$Indian Institute of Space Science $\&$ Technology (IIST),
Trivandrum, India.}
\begin{document}
\date{}
\maketitle
\label{firstpage}

\begin{abstract}
We investigate a relativistic fluid jet driven by radiation from a shocked accretion disc around a non-rotating
black hole approximated by Paczy\'nski-Wiita potential.
The sub-Keplerian and Keplerian accretion rates control the shock location and therefore, the
radiation field around the accretion disc. We compute the radiative moments with full special relativistic
transformation. The effect of a fraction of radiation absorbed by the black hole
has been approximated, over and above the special relativistic transformations. We show that the radiative moments
around a super massive black hole are different compared to that around a stellar mass black hole.
We show that the terminal speed of jets increases with the mass accretion rates,synchrotron emission of the
accretion disc and reduction of proton fraction of the flow composition. To obtain relativistic terminal velocities
of jets, both thermal and radiative driving are important.
We show
for very high accretion rates and pair dominated flow, jets around super massive black holes are truly ultra-relativistic,
while for jets around stellar mass black holes, terminal Lorentz factor of about $10$ is achievable.
\end{abstract}


\section {Introduction}
Astrophysical jets are ubiquitous, as they are associated with
many classes of astrophysical objects such as active galactic
nuclei (AGN e.g., M87), young stellar 
objects (YSO e.g., HH 30, HH 34), X-ray binaries ({\eg} SS433, Cyg
 X-3, GRS 1915+105, GRO 1655-40) etc. 
However, only jets around X-ray 
binaries like GRS1915+105 \citep{mr94} and
AGN like 3C273, 3C345 \citep{zcu95}, M87 \citep{b93} etc are
relativistic. In this paper we concentrate on relativistic jets.
Since it is conjectured
that stellar mass black hole reside at the heart of
microquasars and those of the super-massive variety
dictates the dynamics of the AGNs/quasars, therefore it implies
that a jet has to originate from the accreting matter itself,
since black holes (hereafter, BH) have neither hard surface nor any atmosphere.
Interestingly, simultaneous radio and X-ray observations of microquasars show a very strong
correlation between
the spectral states of the accretion disc and the associated jet
states \citep{gfp03,fgr10,rsfp10}, which reaffirms the fact that jets do originate from
the accretion disc.
In addition, recent observations have shown that jets originate from a region which 
is less than 100 Schwarzschild radii ($r_{\rm g}$)
around the unresolved central object \citep{jbl99,detal12}, which imply 
that the entire disc may not participate
in production of jets, but only the central region of the disc is responsible.

Since jets are supposed to originate very close to the central
object, the plasma at the base should be hot
and is expected to be fully ionized.
This hot outflowing plasma will also be in the intense radiation
field from the accretion disc.
A number of scientists have studied the
interaction of radiation with jets. 
\citet{i80} ignored 
radiation drag and
investigated particle and gas flow in the radiation field of an 
underlying Keplerian disc \citep{ss73}. 
\citet{sw81} studied the interaction of the particle jets with the radiation
field, in the funnel like
region of a thick accretion disc \citep{pw80}. 
Ignoring gravity and for normal electron-proton or $\ep$ plasma jets, the authors achieved
a terminal speed of around $v_T\sim 0.4c$, and obtained
terminal Lorentz factor $\gamma_T \sim 3$ for electron-positron
or, $\el$ plasma.
In a seminal paper, \citet{i89} showed that for particle jets above
 an infinite Keplerian disc, 
the radiation drag
ensures an upper limit of terminal speed, which the author termed it
 as `magic speed' and which turned out to be
$v_{\rm magic}=0.45c$. The much `vaunted' magic speed is actually
 the so-called equilibrium speed
of jet plowing through a radiation field near the Keplerian disc surface. It may be noted that, equilibrium speed
($v_{\rm eq}$)
is the speed of the jet at which the radiation force becomes zero and for speeds above which,
radiative deceleration sets in. 
Equilibrium
speed arises due to the presence of the radiation drag.
Radiation drag is significant for a radiation field due to an extended source. 
On the other hand, terminal speed ($v_T$) is the speed of the jet at which the total
force on the jet approaches zero. Therefore, the jet may achieve $v_T$ only at large distances
away from the central object, while it may reach $v_{\rm eq}$ at distances much closer to the central object.
\citet{mk89}
considered jets which are accelerated
to ultra-relativistic speed way above the local $v_{\rm eq}$. This causes the radiation drag to become 
effective
and decelerate the jet to terminal velocity $v_T \sim 
0.995$. 
\citet{ssbm96} concluded that for the disc they chose, the
maximum
possible of Lorentz factor is $\leq 4$. The Japanese group led by 
Jun Fukue
made very important contribution to the research of transonic outflows, both in the
relativistic, as well as, in non-relativistic domain. \citet{f96} extended Icke's work, and 
studied particle jets away from the axis,
although like Icke, considered near disc approximation for the 
radiation field above it. The actual
terminal jet velocity achieved was $v_T \lsim  v_{\rm magic}$, which 
is expected. However, the main problem
is the tendency of the radiation field to spin up the jet and thereby spreading the jet.
\citet{f99} then studied jets confined by disc 
corona in order to arrest the spreading
and collimate the jet. \citet{hf01} on the other hand, computed
the radiation field due to a Keplerian
disc governed by Newtonian gravity, Schwarzschild gravity and
Kerr gravity. The strength of the
radiation field above the disc,
described by Schwarzschild gravity is $50\%$ lesser than that
due to Newtonian gravity. But the strength of
the radiation field above a disc governed by a Kerr type gravity
is $10$ times higher than that due to the
Newtonian gravity, making radiatively driven jets easier to blow 
for a rotating black hole. 
In a very interesting paper, \citet{fth01} considered a hybrid disc,
consisting of outer Keplerian disc and inner
advection dominated accretion flow or ADAF \citep{nkh97} type 
accretion solutions. Since ADAF is dimmer, so the inner
region from which the electron-positron jet is assumed to
emerge in this paper,
does not contribute in the radiation field. Such a scenario do 
produce jets with terminal Lorentz factor
$\gamma_T \sim 2$, and the radiation field from the outer 
Keplerian disc also helps in collimation.

Along with the various disc models like Keplerian disc, thick
disc, ADAF, investigations
on advective discs was initiated by \citet{lt80,f87,c89}. Such a
disc can admit smooth solutions,
and for the right choice of parameters it may harbour shock transition. This 
disc model was extended by considering injection
of a mixture of matter with Keplerian angular momentum and 
sub-Keplerian angular momentum. The portion of disc which is termed as sub-Keplerian
disc (SKD), sandwiches the 
Keplerian disc (KD) from the top and the bottom (see Fig. 1).
SKD may undergo shock transition and
at the shock, KD terminates due
to extra heating in the post-shock disc or PSD \citep{ct95}. Although
it was initially proposed
as an elaborate and contrived model solution, but was recently 
confirmed by numerical simulation
\citep{gc13}. Interaction of radiations with the outflowing jet 
from PSD
was studied by \citet{cc00a,cc00b,cc02}. The investigation of 
outflowing jet \citep{cdc04,c05}
was further elaborated
for the radiation field of the Chakrabarti-Titarchuk type hybrid
disc \citep{ct95}. Generally 
most of the papers which investigate the interaction of jet with
disc photons
do not consider the issue of the launch mechanism of 
jets in terms of the accretion properties.
In the advective disc regime, numerical simulations
first showed that the extra-thermal gradient force in the PSD
automatically generates bipolar outflows \citep{mlc94,mrc96,dcnm14}. 
Theoretically too, for
a viscous advective disc, the mass outflow rate of bipolar
outflow was computed in terms of various accretion disc parameters
\citep{cd07,kc13,kscc13}.
However, jets emerging from the PSD
in the steady state and hydrodynamic limit are weak. On the other hand, if the emanating
jets
are simultaneously acted on by disc photons, then the jets obtained
are stronger \citep{kcm14}. In fact, it has been shown by 
\citet{kcm14}, that as the advective
disc spectral state moves from low-hard or LH state to intermediate
states or IM, the steady jet becomes stronger, as has been reported in observations
\citep{gfp03,fgr10,rsfp10}. In other words, \citet{kcm14} not only
generated jets from accretion solutions,
but also accelerated the jets by depositing momentum of disc 
photons on to the jet.
However, the formalism followed by \citet{kcm14} is only correct up to the first order of 
$v/c$, where $v$ is the flow velocity. So solving the jet equations
in the relativistic hydrodynamic limit is warranted for Chakrabarti-Titarchuk type disc.

The equations of motion of radiation hydrodynamics were developed 
by many authors \citep{hs76,mm84,kfm98}.
It was observed that in the non-relativistic limit, only flux of the
radiation pushes matter. In the relativistic limit
and optically thin plasma, the moving plasma is pushed by the flux and
is dragged by the energy density and pressure of the radiation field.
The radiation transfers 
momentum on to the electrons and therefore, the radiative acceleration term is proportional to the number density
of electrons/positrons in the jet, which is true in relativistic, as well as, non-relativistic limit.
But in relativistic limit, the radiative acceleration term is also proportional to 
the inverse of the enthalpy of the flow, which marks a major difference between relativistic and non-relativistic
domain.
Therefore, although \citet{kcm14} showed radiatively driven
jets qualitatively explain the correlation between jet states
and the spectral state of the accretion discs, but a correct
relativistic narrative is required.
If however, one considers the gas pressure ($p$) to be negligible compared to
the radiation pressure ($P^{ij}$), then even in the fully relativistic limit,
the net radiative term is only proportional to the number density of the flow.
Therefore, the physics of interaction of
disc radiation with a fluid jet in the relativistic regime is 
qualitatively and quantitatively different than
that between radiation and jet in the non-relativistic domain \citep{cc02,kcm14}, or relativistic
jets in the domain where $P_{ij} \gg p\delta_{ij}$ \citep{cdc04,c05}. 
In this paper, we would
like to investigate this phenomenon in details. \citet{ftrt85}
studied interaction of fluid jets with the disc radiation
but the jet fluid was described by fixed adiabatic index $\Gamma$ and Newtonian gravitational potential.
However,
fixed $\Gamma$ is an artifact of non-relativistic kinetic theory, {\ie}
if the internal random motions of the particles
in the fluid is relativistic, then $\Gamma$ is a function of 
temperature \citep{c38,s57}, where,
for relativistic temperatures, the flow is described by 
$\Gamma \sim 4/3$ and for non-relativistic
temperatures $\Gamma = 5/3$. \citet{t48} showed that
it is unphysical to consider a fixed $\Gamma$ to describe 
fluid with several orders of magnitude variation in temperature.
Moreover, it has also been shown that, not only temperature, 
relativistic nature of the thermal energy depends also on the
composition of the flow \citep{c08,cr09,cc11,kscc13,crj13,kc14}.
Infact it was shown that, contrary to expectation,
pair-plasma is thermally the least relativistic, and to make a 
flow more relativistic one needs baryons in addition 
to electrons or positrons, and the fluid with proton number 
density around $27\%$ of that of electrons is thermally the
most relativistic. Now radiation force is imparted mainly 
onto the electrons and positrons which should make the
lighter jet to move faster due to inertia, as is shown for 
cold jets \citep{cdc04,c05}.
On the contrary, the thermodynamics
of the jet will make the fluid with $27\%$ proton content to
be more relativistic. So in presence of radiation driving, would
a jet with protons less than $27\%$ be more accelerated, or, the thermal nature
of the flow dictate the dynamics. 
Additionally for fluid jets, the pressure gradient force should accelerate the jet,
while gravity decelerate, and radiative force depends on the fluid
speed in addition to the radiative moments.
At the base, the jet is supposed to be hot, so initially, can the jet be thermally driven to speeds
above $v_{\rm eq}$, and therefore, is mostly decelerated by radiation and not accelerated at all?
In short, can disc radiation power jets to relativistic terminal speeds?
These are few of the various issues we would like to investigate in this paper.
In this paper the accretion disc plays an auxiliary role, {\ie} we do not employ a self consistent
calculation of accretion-ejection system, rather the jet is assumed to originate in the inner
disc, which in our case is PSD, and the accretion disc just provides the radiation field
which interacts with the jet. While doing so, we have not increased the number of accretion disc parameters,
and the shock location is estimated from the SKD accretion rate, and PSD luminosity has been estimated
from the accretion rates from SKD and KD.

In the next section, we present the simplifying assumptions and various equations.
In section \ref{sbsec2.1}, we present the governing equations
of the jet. In section \ref{sbsec2.2}, we present the model of accretion disc, estimation of the disc intensity
from various disc components (section \ref{sbsbsec2.2.1}), and computation of the radiative moments
(section \ref{sbsbsec2.2.2}) from the disc intensities.
Disc intensities, the disc velocity and temperature profiles are estimated in
Appendix \ref{app:accvT}. The post-shock intensity is estimated in Appendix \ref{app:mdotxs}.
In section \ref{sec3}, we present the solution methodology and in section \ref{sec4}, we present the results and in
the last section (\ref{sec5}) we discuss the physical implications of the obtained 
results and draw concluding remarks.

\begin {figure}[h]
\begin{center}
 \includegraphics[width=16.cm]{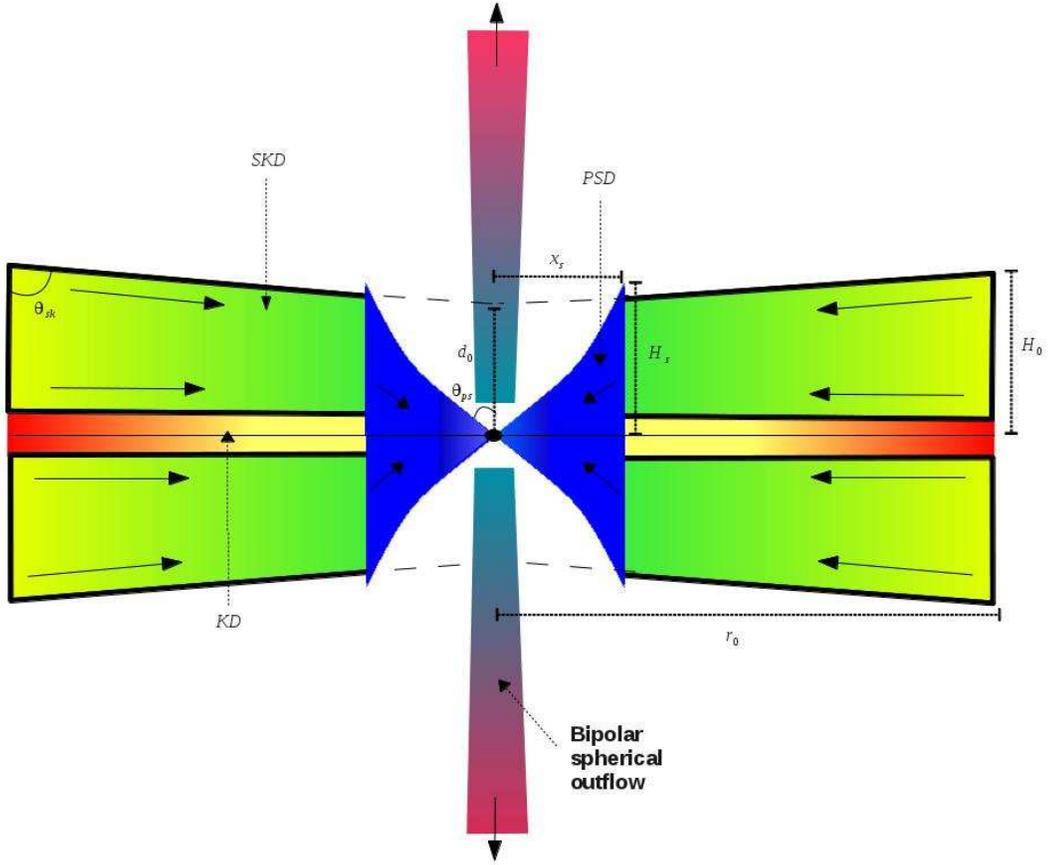}
\vskip -0.5cm
 \caption{Cartoon diagram of cross-sections of axis-symmetric
accretion disc and the associated jet in ($r,~\phi, z$ coordinates). The Keplerian disc (KD),
sub-Keplerian disc (SKD), and the post-shock disc (PSD) are broadly the three regions of the disc.
The shock location
$x_s$, the intercept of SKD on the axis ($d_0$), height of the shock $H_s$ are shown, the outer edge of the
disc $r_0$ are all marked. Semi-vertical
angle of PSD is $\theta_\s$ and for SKD it is $\theta_\sk$. The gradient
of colour represents low temperature (red) to high temperature (blue).}
\label{lab:geom}
 \end{center}
\end{figure}

\section{Assumptions, governing equations of jet \& structure of accretion disc}
\label{sec2}
 The space time metric is given by
\be
ds^2=-c^2dt^2+dr^2+r^2d\phi^2+dz^2
\ee 
where, $t,~r,~\phi,~z$ are the time, radial, azimuthal and the axial coordinates.
The jet is considered to be in steady state ({\ie} $\partial/\partial t=0$).
The above is a special relativistic metric.
Although the jets originate close to the central compact object, but it traverses
to distances where the effect of gravity is negligible. Hence pseudo potential is used
to take care of gravity in the
equations of motion, such that at close ranges the gravity limits the outward thrust,
and at large distances it is just special relativistic regime, therefore
one may avoid general relativistic complications
and still be accurate enough. Moreover, the jets are collimated, so we
consider on-axis ({\ie} $u^r=u^{\phi}=\partial/\partial r=0$) and axis-symmetric ($\partial/\partial \phi=0$) jet.
Without any loss of generality, the jet is assumed to expand radially along the $z$-axis,
the value of any jet variable on any particular point of the axis, is assumed to be maintained along its
breadth at the same $z$.
In the present effort,
we just assume that there are jets from an accretion disc, but do not compute jets self
consistently from accretion solution. The accretion disc is present but plays a supportive role,
by supplying the radiation which drives the jet. In the next subsection we present the
equations of motion. We present
the description of the accretion disc
and the method to compute radiative moments at section \ref{sbsec2.2}.

\subsection {Equations of motion of the jet}
\label{sbsec2.1}
The energy momentum tensor of the jet and the radiation field is given by
\begin{equation}
T^{\alpha \beta}_M=(e+p)u^{\alpha}u^{\beta}+pg^{\alpha \beta};
~~T^{\alpha \beta}_R={\int}I_{\nu}l^{\alpha}l^{\beta}d{\nu}d{\Omega},
\end{equation}
where, suffix $M$ stands for jet material and $R$ stands for radiation field. The internal energy
of the jet is $e$ and $p$ is the isotropic pressure of the jet fluid, the metric
tensor components are given by $g^{\alpha \beta}$ and $u^{\alpha}$ is the four velocity.
Furthermore, $I_{\nu}$ is the specific intensity
of the radiation field, $l^{\alpha}$s are the directional derivatives, $\nu$ the frequency of the radiation
and $\Omega$ is the solid angle subtended by the field point on to the source point.
By definition, field point is where the moments are computed, while source point
is the location of the source of the radiation. 
The equations of motion are given by
\begin{equation}
T^{\alpha \beta}_{;\beta}=0=(T^{\alpha \beta}_{R}+T^{\alpha \beta}_{M})_{;\beta}
\end{equation}
From the above equation, the momentum conservation equation, in the $i^{th}$ 
direction is obtained by using projection tensor $(g^{i}_{\alpha}+u^iu_{\alpha})$,
{\ie}
\begin{equation}
(g^{i}_{\alpha}+u^iu_\alpha)T^{\alpha \beta}_{M_{;\beta}}=-(g^{i}_{\alpha}+u{^iu_\alpha})
T^{\alpha \beta}_{R_{;\beta}}
\label{genmomb.eq}
\end{equation}
Similarly, the energy conservation equation is obtained by taking 
\begin{equation}
u_{\alpha}T^{\alpha \beta}_{M_{;\beta}}=-u_{\alpha}T^{\alpha \beta}_{R_{;\beta}}
\label{genfstlaw.eq}
\end{equation}
The derivation of the equations of motion of radiation hydrodynamics
for optically thin plasma, using above preliminaries, was 
investigated by a number of workers. 
Since we study on axis jet, the equations
of motion greatly simplifies. 
The momentum balance equation (equation \ref{genmomb.eq}), in steady state and for on axis jet becomes;
\be
(e+p)\left(u^z\frac{du^z}{dz}+\frac{GM_{B}}{(z-r_g)^2} \right)
= -\frac{dp}{dz}-u^zu^z
\frac{dp}{dz}+{\rho}_e\frac{{\sigma}_{T}
}{m_ec}{\Im}^z.
\label{mom.eq}
\ee
The term containing $M_B$ in the r. h. s of equation (\ref{mom.eq}) is the Paczy\'nski-Wiita
term mimicking the gravity of non-rotating BH \citep{pw80}.
The energy conservation equation (\ref{genfstlaw.eq}) in the scattering regime is,
\begin{equation}
\frac{de}{dz}-\frac{e+p}{\rho}\frac{d\rho}{dz}=0
\label{econ.eq}
\end{equation}
where, $\rho$ is the total mass density, $\rho_e$ is the leptonic mass density of the flow and $m_e$ is the electron rest
mass.
Similarly, from continuity equation the mass outflow rate is given as 
\begin {equation}
\dot {M}_{out}=\rho u^z {\cal A},~~ {\cal A} \propto z^2 \mbox{ for radial, narrow jet about the axis!}
\label{mdotout.eq}
\end {equation}

In above equations, $G$, $M_{B}$, ${\sigma}_{T}$,
 $r_g=2GM_B/c^2$ and ${\cal A}$
are the universal gravitational constant, the mass of the central
 black hole, Thomson scattering cross-section,
Schwarzschild radius and cross section of the jet respectively.
${\Im}^z$ is the net radiative contribution and is given by;
\begin{eqnarray}
\hskip 1.0cm
\frac{{\sigma}_T}{m}\frac{{\Im}^z}{c} & = & \frac{{\sigma}_T}{m_e}\left[
{\gamma}\frac{F^z}{c}-{\gamma}^2u^zE
-u_{j}P^{zj}+u^z
\left(2\frac{{\gamma}}{c}u_{j}F^{j}-u_{j}u_{k}P^{jk}
\right) \right] \\ \nonumber
& = &  [\gamma(1+2u^zu_z){\cal F}-\gamma^2u^z{\cal E}-{\cal P}(u_z+u^zu_zu_z)]~~\mbox{for on axis jet! }
\label{fz.eq}
\end{eqnarray}

In above equations, $E$, $F^z$, and $P^{zz}$
are the radiative energy density, the radiative flux
and the radiative pressure tensor measured in observer
frame, and
${\cal E}=\frac{{\sigma}_T}{m_e}E$, ${\cal F}=\frac{{\sigma}_T}{m_ec}F^z$,
and ${\cal P}=\frac{{\sigma}_T}{m_e}P^{zz}$.
Furthermore, ${\gamma}~(\equiv -u_t={\sqrt{1+u_iu^i}})$ is the Lorentz factor.

It may be noted that, we have assumed the jet to be flowing radially out within a conical surface for simplicity,
since our primary concern is to investigate the respective role played by thermal and radiative driving terms
on jet dynamics. Needless to say depending upon initial
condition and disc radiation field, the jet geometry may depart from the simple geometry we are following in this paper.
Even then, our assumption is not completely outlandish. The funnel like surface of the PSD is the region
from where the jet is supposed to originate, the shape itself will arrest the lateral spread
of the outflowing matter. Moreover, such shape causes the $r$ component of
radiative flux directed towards the axis \citep{c05} which would also reduce the lateral spreading
even with high jet-base temperature.
However, one can justify our assumption only if we solve the jet equations since we need some estimate of the pressure.
In appendix (\ref{app:jgeom}), we estimated the pressure gradient term along with the radiative term along $r$ and compared
them, the assumption of conical jet cross-section seems to hold.

\subsubsection {Equation of state and the final form of equations of motion}
\label{sbsbsec2.1.1}
The physics of the jet propagating in the radiation field of the accretion disc
can be understood, if equations (\ref{mom.eq}-\ref{mdotout.eq}) are simultaneously solved.
However, one has also to supply a closure relation {\ie} a relation between $e,~p,~\rho$
called the equation of state (EoS) in order to solve equations (\ref{mom.eq}-\ref{mdotout.eq}).
An EoS for multispecies, relativistic
flow proposed by \citet{c08,cr09} is adopted, and is given by,
\begin{equation}
e=n_{e^-}m_ec^2f
\label{eos.eq}
\end{equation}
with $n_{e^{-}}$ is the electron density and $f$ is given by
\begin{equation}
f=(2-\xi)\left[1+\Theta\left(\frac{9\Theta+3}{3\Theta+2}\right)\right]
+\xi\left[\frac{1}{\eta}+\Theta\left(\frac{9\Theta+3/\eta}{3\Theta+2/\eta}
\right)\right]
\label{eos2.eq}
\end{equation}
Here, non-dimensional temperature is defined as
$\Theta=kT/(m_ec^2)$, $k$ is the Boltzmann constant and
$\xi = n_{p^{+}}/n_{e^{-}}$ is the relative proportion of protons with respect to the number density of electrons.
The mass ratio of electron and proton is $\eta = m_{e}/ m_{p^{+}}$. It is easy to see that
by putting $\xi=0$, we generate EoS for relativistic $\el$ plasma \citep{rcc06}.
The expressions of the polytropic index $N$, adiabatic index $\Gamma$ and
adiabatic sound speed $a$ are given by
\begin{equation}
N=\frac{1}{2}\frac{df}{d\Theta} ;~~ \Gamma=1+\frac{1}{N} ; ~~
\frac{a^2}{c^2}=\frac{\Gamma p}{e+p}=\frac{2 \Gamma \Theta}
{f+2\Theta}.
\label{sound.eq}
\end{equation}
This EoS is an approximated one, and the comparison with the exact one shows that this EoS
is very accurate (appendix \ref{app:eos}). Additionally, being algebraic and avoiding the presence
of complicated special functions,
this EoS is very easy to be
implemented in simulation codes, as well as, be used in analytical investigations \citep{cr09,cc11,rcc06,crj13}.
The jet plasma is fully ionized.
Therefore the interaction with
photons would be dominated by scattering.
Therefore, the energy equation (\ref{econ.eq}) has no source term
because in the scattering regime and in absence of emission/absorption,
the r. h. s is zero
and the flow is isentropic \citep{mm84}. Under such conditions, equation (\ref{econ.eq}) along with equation
(\ref{eos.eq}) can be integrated
to obtain the relativistic isentropic equation of state,
$$
\rho={\cal C}\mbox{exp}(k_3) \Theta^{3/2}(3\Theta+2)^{k_1}
(3\Theta+2/\eta)^{k_2},
$$
where, $k_1=3(2-\xi)/4$, $k_2=3\xi/4$, $k_3=(f-\tau)/(2\Theta)$, $\tau=(2-\xi+\xi/\eta)$ and ${\cal C}$ is the constant
of entropy.
We replace $\rho$ from the above equation on to equation (\ref{mdotout.eq}), we get the expression for entropy-accretion
rate,
\begin{equation}
{\dot {\cal M}}_{out}=\frac{{\dot M}_{\rm out}}{{\rm geom. const.}{\cal C}}=\mbox{exp}(k_3) \Theta^{3/2}(3\Theta+2)
^{k_1}
(3\Theta+2/\eta)^{k_2}u^zz^2
\label{entacc.eq}
\end{equation}
This is also a measure of entropy of the jet and remains constant 
along the jet.
We adopt a unit system where, the unit of speed is $c$,
unit of length $r_g=2GM_B/c^2$ and the unit of mass is $M_B$.
Henceforth we write all equations in this unit system, except where it is explicitly mentioned.
The three-velocity $v$ is given by $v^2=-u_iu^i/u_tu^t=-u_zu^z/u_tu^t$, {\ie}
$u^z=u_z={\gamma}v$.
Now using energy conservation equation (\ref{econ.eq}) along 
with the equation of state (\ref{eos.eq}), the gradient of 
temperature of jet is given by,
\begin{equation}
\frac{d{\Theta}}{dz}=-\frac{{\Theta}}{N}\left[ \frac{{\gamma}
^2}{v}\left(\frac{dv}{dz}\right)+\frac{2}{z}\right]
\label{dthdr.eq}
\end{equation}
The momentum balance equation (\ref{mom.eq}), with the help of equations (\ref{eos.eq}), (\ref{sound.eq}) and
(\ref{dthdr.eq}), becomes 
\begin{eqnarray*}
{\gamma}^4v\left(1-\frac{a^2}{v^2}\right)\frac{dv}
{dz} & = & \frac{2{\gamma}^2a^2}{z}-\frac{1}{2(z-1)^2} \\ \nonumber
& + & \frac
{{\gamma}^3{(2-\xi)}}{f+2{\Theta}}[(1+v^2){\cal F}-v
({\cal E}+{\cal P})] 
\end{eqnarray*}
\be
~~~~~~~~~~~~~~~~~~~~~~ = {\rm a}_t+{\rm a}_g+{\rm a}_r.
\label{dvdr.eq}
\ee
The l. h. s is the net acceleration term of a steady state jet. On the r. h. s, the first term is the thermal term a$_t
=2{\gamma}^2a^2/z$
and it accelerates, while the second being gravity a$_g=-0.5/(z-1)^2$, it decelerates. The third term in r.h. s is the radiative term
a$_r={\gamma}^3{\tau}[(1+v^2){\cal F}-v
({\cal E}+{\cal P})]/(f+2{\Theta})$. The radiative contribution is within the square bracket and the rest represents the interaction of
matter jet with the radiation field.
The physical significance of the term in the square bracket term is worth noticing.
It has the form
$$
(1+v^2){\cal F}-v({\cal E}+{\cal P})
$$
The term proportional to $v$ comes with a negative sign and would decelerate and is called the radiation drag term.
If the first term
$(1+v^2){\cal F}$ dominates, then radiation would accelerate the flow, which means the net radiative
term would either be accelerating or decelerating depending on the velocity. The dependence of radiative term
on $v$ arises purely due to relativity. In the purely non-relativistic domain {\ie} $v\ll 1$, the radiative term
is just ${\cal F}$. In the fast but sub relativistic domain {\ie} $v^2 \ll 1$ the radiative term is
${\cal F}-v({\cal E}+{\cal P})$ similar to the formalism followed by \citet{cc02,kcm14}. The drag term arises
due to the resistance faced by the moving material through the radiation field, and the finite value of the speed of light.
Much talked about equilibrium speed $v_{\rm eq}$ is when ${\rm a}_r=0$, {\ie}
\be
v_{\rm eq}=\Re-\sqrt{\Re-1};~\mbox{where, } \Re=\frac{{\cal E}+{\cal P}}{2{\cal F}}.
\label{veq.eq}
\ee
From equation (\ref{veq.eq}), it is clear that if the relative contribution of radiative moments
or $\Re$ approaches $1$, {\ie} ${\cal F}={\cal E}={\cal P}$, then $v_{\rm eq} \rightarrow 1$, {\ie} no radiation drag.
Therefore, the nature of the quantity $\Re$ dictates, whether a radiation field
would accelerate a flow or decelerate it.
Of course the resultant acceleration depends on the magnitude
of all moments. There is an added feature of radiatively driven
relativistic fluid, {\ie} the radiative term is multiplied by a term inverse
of enthalpy ($\{f+2\Theta\}/\tau$) of the flow, which actually suggests
that the effect of radiation on the jet is less for hotter flow.

\subsection{Accretion disc and radiative moments}
\label{sbsec2.2}
The accretion disc model considered here is the hybrid disc of the Chakrabarti-Titarchuk flavour
\citep{ct95,gc13}. In Fig. \ref{lab:geom}, we show all the components of the disc, {\ie} PSD, SKD and KD.
The SKD flanks the KD, but mingles and forms the single component PSD.
The colour coding represents lower (red) to higher (blue) temperature. The outer edge of the disc is $r_0$
where the disc height is $H_0$. The inner edge of SKD and KD is the shock location $x_s$, the inner edge of PSD
say $r_{\rm in}$ is in principle the horizon, but we have considered it to be $r_{\rm in}=1.5r_g$ while calculating the
radiative moments, since very little radiation is expected from a region very close to the horizon.
The shock height is marked as $H_s=0.6~(x_s-1)$ \citep{ct95,c05}, therefore $\theta_\s={\rm tan}^{-1}(x_s/H_s)$.
The semi vertical angle of the SKD
($\theta_\sk$ in Fig \ref{lab:geom}) is taken to be $85^\circ$. This assumption has been dictated by
a large number of simulations, which showed SKD to have a flatter surface compared to PSD \citep{mlc94,mrc96,gc13,dcnm14}.
The intercept of the SKD surface on the z-axis $d_0=0.4\times H_s$.
We chose $r_{0}$=3500$r_g$.
SKD emits via synchrotron and bremsstrahlung processes, so the information of velocity ({\ie} density) and temperature
profile is required, and have been estimated in appendix A.
Injection speed ($\vartheta_0$) for the SKD at the outer disc boundary is kept $0.001$,
the angular momentum of the disc $\lambda$ is 1.7 and the 
temperature at $r=r_0$ is $\Theta_0=0.1$.
The PSD
is hotter than the rest of the disc (including SKD and KD) and puffs up in the form of a torus.
KD emits thermal photons \citep{ss73} and SKD emits via bremsstrahlung and synchrotron processes.
PSD emits bremsstrahlung and synchrotron photons, as well as, being fatter and hotter, inverse Comptonize these photons
and the photons intercepted from SKD and KD to produce hard radiation.
All the spectral states therefore, can be obtained by controlling the SKD accretion rate 
${\dot M}_\sk$ and the KD accretion rate ${\dot M}_\kd$. If ${\dot M}_\kd$ is relatively less than ${\dot M}_\sk$,
then due to the lack of supply of soft photons, PSD will remain hot, and thus producing
the low hard or LH state. Increase in viscosity and/or increase in
accretion rate at the outer boundary would push the shock closer to the central object
\citep{kc13,kc14}. This would brighten up the disc, but would make the spectra softer as the number of
hard photons from the PSD would be lower due to the decrease in PSD size. Hence the spectral index would
increase and spectra would enter the intermediate or IM states. The increased size of the pre-shock disc
(SKD+KD) and the increased accretion rate, would eventually weaken the PSD or completely destroy it,
the contribution of hard photons would plummet and the disc would be more luminous, while the spectra
would become soft similar to a 
multicoloured black body.

\subsubsection{Relativistic transformations of intensities from various disc components}
\label{sbsbsec2.2.1}
In order to compute radiative moments, we need to know radiative intensities
of various disc components.
The intrinsic KD intensity is given by \citep{ss73} 
\begin{equation}
I_{\kd 0}=\frac{3GM_{B}{\dot M}_\kd}{8{\pi}^2r^3}\left(1-
{\sqrt{\frac{3r_g}{r}}} \right) \ \  {\rm erg} \
{\rm cm}^{-2} {\rm s}^{-1}
\label{kdint.eq}
\end{equation}

To compute the radiative moments from SKD, we need to know the temperature and density distribution
of SKD, in order to calculate the intrinsic radiative intensity of SKD.
The density and temperature of SKD starting with some outer boundary condition can only be solved numerically.
However for simplicity,
we estimate 
the approximate values of velocity, density and temperature profile of SKD in order to compute the $I_{\sk 0}$.
Equations (\ref{lofac.eq}, \ref{velsk.eq}) give the analytical expression of all the components of
three-velocities and the corresponding Lorentz factors (Appendix \ref{app:accvT}) of the SKD. This
allows us to compute the density profile for a given ${\dot M}_\sk$.
The density and temperature
profile of SKD are estimated in equations (\ref{densrat.eq} \& \ref{thetacc.eq}) in Appendix  \ref{app:accvT2}.

We further assume that there is stochastic magnetic field in the SKD which is in partial equipartition with the gas pressure.
The ratio of magnetic pressure ($p_{\rm mag}$)
and the gas pressure ($p_{\rm gas}$) is also assumed to be constant $\beta$
{\ie} $p_{\rm mag}=B^2/8\pi=\beta p_{\rm gas}=\beta n_\sk k T_\sk$, where, $n\sk$ and $T_\sk$ are the SKD
local number density and temperature, respectively.
The emission mechanism is dominated by synchrotron and bremsstrahlung emission,
and therefore the SKD intensity is given by \citep{s82,st83,kcm14,kc14}
$$
I_{\sk_0}=I_{\rm syn}+I_{\rm brem}
$$
\begin{equation}
=\left[\frac{16}{3}\frac{e^2}{c}\left( \frac{eB_\sk}{m_e c} \right)^2
 \Theta^2_\sk n_{\sk} r+ 1.4\times 10^{-27}n_{\sk}^2g_bc \sqrt{\frac{\Theta_\sk 
m_e}{k}}\right]\frac{\left(d_0~sin \theta_{\sk}+r~cos \theta_{\sk}
\right)}{3} \ \  {\rm erg} \
{\rm cm}^{-2} {\rm s}^{-1}
\label{skint.eq}
\end{equation}
where, $\Theta_\sk, n_\sk, r$, $\theta_\sk$, $d_0$, $B_\sk$ and $g_b$ are the 
pre-shock local dimensionless temperature,
electron number
density, horizontal distance from center of the disc, angle from
the axis of symmetry to the pre-shock surface, the intercept of the SKD surface on to the axis of symmetry, 
the magnetic field and relativistic Gaunt factor $(g_b=1+1.78\Theta^{1.34})$, respectively.
The factor outside square brackets
serves as the conversion factor from emissivity (${\rm erg}. 
{\rm cm}^{-3}{s}^{-1}$) into intensity  (${\rm erg}. {\rm cm}^
{-2}{s}^{-1}$).

The expression of intensity is more complicated for PSD, as a result we make further simplifying assumptions.
The PSD itself emits via bremsstrahlung and synchrotron, but also inverse Comptonizes its own photons,
as well as, photons intercepted from SKD and KD. It is beyond the scope of this paper to do a proper
radiative transfer treatment of the accretion disc. Instead, we fed the code of \citet{mc08} with the
viscous and dissipative solutions of \citet{kc14} as back ground solution, and computed the spectra
from PSD, SKD and KD. The shock location $x_s$ is estimated from equation \ref{xsdotm.eq} and is
also presented in Fig. \ref{lab:FigB1}(a). Although the relation between $x_s$ and ${\dot M}_\sk$
has been obtained by generalizing the solutions for a certain viscosity
parameter \citep{kc14}, but we treat them as generic, since the general pattern is similar for a large number
of cases we have analyzed. Then we fit the ratio of the luminosities of PSD and that
from the pre-shock disc (SKD and KD) $\chi$, as a function of $x_s$ from the spectra obtained by the radiative transfer of the
background solutions. The relation between $\chi$ and $x_s$ is given by equations \ref{chi1.eq}, \ref{chi8.eq} and is
also plotted in  Fig. \ref{lab:FigB1}(b).
The preshock luminosities are obtained by integrating $I_\sk$ and $I_\kd$
over the respective disc surfaces.
So in principle we have two free parameters to fix the radiation field above
the accretion disc, ${\dot M}_\sk$ and ${\dot M}_\kd$.
The intensity as measured in local rest frame of PSD is given by 
\be
I_{\s 0}=L_\s/{\pi}{A_\s}={\ell_\s}L_{\rm Edd}/{\pi}{A_\s}~({\rm erg} {\rm cm}^{-2} {\rm s}^{-1}),
\label{psdint.eq}
\ee 
where $L_\s$ and ${A_\s}$
are the PSD luminosity and the surface area of the PSD
respectively. $L_{\rm Edd}$ is the Eddington luminosity and ${\ell}_\s$ is the PSD
luminosity in units of $L_{\rm Edd}$.

The intensities from either the PSD or SKD or KD components of the disc,
namely, $I_{\s 0},~ I_{\sk 0}$ and $I_{\kd 0}$, are
measured in the local rest frames of the disc components. However, the matter
in the disc is moving, so one has to transform the quantities into the observer frame!
The intensity measured in the observer frame is presented in compact notation as,
\begin{equation}
I_j=\frac{I_{j_0}}{\gamma^4_{j}\left[1+{\rm v}_il^i\right]^4_j}
\label{tran_int.eq}
\end{equation}
Here $\gamma_{j}$ is Lorentz factor
and ${\rm v}^i$ is $i^{th}$ component of 3-velocity of accreting matter and $l^i$s are directional cosines. 
The suffix $j\rightarrow \s,\sk,\kd$ signifies the
contribution from either PSD, or SKD, or KD.
For PSD and SKD ${\rm v}^i$ is calculated following Appendix \ref{app:accvT}, while for KD ${\rm v}^i \equiv
(0,{\rm v}_\kd,0)$
is the Keplerian azimuthal velocity.
The luminosity from various components of the disc is obtained by integrating the respective local specific intensities
over the disc surface {\ie}
\be
L_j=2\int I_j\times 2\pi r~{\rm cosec}^2\theta_j~dr.
\label{lum.eq}
\ee
here, $j$ represents various disc components. To compute the luminosity,
the limits of integration are inner and outer limits of
the disc components, for example for PSD the limit of integration is $r_{\rm in} \rightarrow x_s$, while
for SKD and KD the integration limits are $x_s \rightarrow r_0$. Apart from integration limits,
various disc components are identified by the respective $\theta_j$s, which defines the surface of
disc components.
The total luminosity is given by $L=L_\s+L_\sk+L_\kd$, and in units of Eddington limit it is
$\ell=\ell_\s+\ell_\sk+\ell_\kd$.
All the transformations presented above, are exactly correct in the special relativistic regime.
We are not taking into consideration the phenomenon of light bending since the jet spans from close to the
horizon to few thousands of $r_g$. Beyond few tens of $r_g$ the general relativistic effect may be
ignored, but special relativistic effects cannot be. But close to the horizon, if only
special relativistic effects are considered then the $I_\s$ gets unnecessarily jacked up \citep{c05}.
In order to address this, we estimate the fraction
of radiation that will not be absorbed by the black hole and escape.
From geodesic equations of photons, it can be easily shown that
if ${\rm sin}\psi>3\sqrt{3}(1-1/\varpi)/(2\varpi)$ then the photon escapes \citep{st83}, where
$\psi$ is the angle of the direction of propagation of light with the radial direction,
and the $\varpi$ is the spherical radius coordinate.
So we express $\psi$ in terms of $r$ and $\theta_\s$ the semi 
vertical angle of the PSD inner surface. Assuming locally the radiation is isotropic,
then the fraction of intensity from PSD which would escape and interact with the jet is,

\begin{equation}
{\cal R}=\frac{\pi-sin^{-1}\left(3\sqrt{3}(sin\theta_\s/2r)(1-sin\theta_\s/r)\right)}{\pi}
\label{rat.eq}
\end{equation}
Since SKD and KD is further away from the black hole, no such estimation of gravity effect on
emitted radiation from these components is needed.

\subsubsection {Computation of radiative moments}
\label{sbsbsec2.2.2}
Radiative moments are the zeroth, first and second moments of 
specific intensity, and frequency integrated moments are respectively called radiation 
energy density, radiative flux and radiation pressure, and are expressed as following,
$$
{
\ \left( \begin{array}{c}
E\\
F^i\\
P^{jk}\end{array}\right)=\int \int{
\ \left( \begin{array}{c}
\frac{1}{c}\\
l^i\\
\frac{l^j l^k}{c}\end{array}\right)I_{\nu}}d\nu d\Omega}
$$
Here, $l^i$s are directional derivatives, $\nu$ the frequency of radiation and $\Omega$ is the
solid angle subtended by the field point (where the moments are calculated) on to the source of
radiation. Since we are considering on axis jet, therefore we need to compute the radiative moments
only along the axis.
The radiative moments along the jet axis ($z$ axis) are calculated
from the PSD, SKD and KD components of this hybrid disc model, these components
are indicated in following expressions with subscripts $\s$, $\sk$ 
and $\kd$ respectively.

\begin{eqnarray}\nonumber
 {\cal E}=\frac{\sigma_T}{mc}\left(\int I_\s d\Omega_\s + 
\int I_{\sk} d \Omega_\sk + \int I_\kd d \Omega_\kd \right)=
\frac{\sigma_T}{m}\left(E_\s+E_\sk +E_\kd\right)
\\ ={\cal E}_\s+{\cal E}_\sk+{\cal E}_\kd
\label{ee1.eq}
\end{eqnarray}
\begin{eqnarray}\nonumber
 {\cal F}=\frac{\sigma_T}{mc}\left(\int I_\s l^z 
d\Omega_\s + \int I_{\sk}l^z
d \Omega_{\sk} + \int I_{\kd}{l^z}
d \Omega_{\kd}\right)=\frac{\sigma_T}{mc}\left(F_\s+F_\sk
+F_\kd\right)
\\
={\cal F}_\s+{\cal F}_\sk+{\cal F}_\kd
\label{ff1.eq}
\end{eqnarray}
\begin{eqnarray}\nonumber
 {\cal P}=\frac{\sigma_T}{mc}\left(\int I_\s l^zl^z
d\Omega_\s + \int I_{\sk}l^zl^z
d \Omega_{\sk}+ \int I_{\kd}l^z l^z
d \Omega_{\kd}\right)=\frac{\sigma_T}{m}\left(P_\s
+P_\sk+P_\kd\right)
\\
={\cal P}_\s+{\cal P}_\sk+{\cal P}_\kd
\label{pp1.eq}
\end{eqnarray}

All the points on the axis of symmetry are field points ($z$), {\ie} where
radiative moments are to be computed. The coordinates on the disc surface
are $r,~\phi,~ z^{\prime}$ and $z^{\prime}_j=r~cot\theta_j$, where $j\equiv \s/\sk/\kd$.
It is easy to see that for extended source, and field point close to the source,
the directional cosines of $l^z<1$, but for point source $l^z=1$, and ${\cal E}={\cal F}={\cal P}$.
Therefore, according to the note following equation (\ref{dvdr.eq}), for point sources
${\Re}=({\cal E}+{\cal P})/2{\cal F}=1$ and there will be no radiation drag.
The radiative moments from PSD are,
\begin{equation}
{\cal E}_{\s}={\cal S}\int^{x_s}_{r_{\rm in}}\int^{2\pi}_{0}
\frac{{\cal R}zrdrd{\phi}}
{[(z-r~cot\theta_\s)^2+r^2]^{3/2}\gamma^4_{\s}\left
[1+{\rm v}_il^i \right]^4_\s}
\label{epsd.eq}
\end{equation}
\begin{equation}
{\cal F}_{\s}={\cal S}\int^{x_s}_{r_{\rm in}}\int^{2\pi}_{0}
\frac{{\cal R}z(z-r~cot\theta_\s)rd{\phi}dr}{[(z-r~cot\theta_\s)
^2+r^2]^2\gamma^4_{\s}\left[1+{\rm v}_il^i \right]^4_\s}
\label{fpsd.eq}
\end{equation}
\begin{equation}
 {\cal P}_{\s}={\cal S}\int^{x_s}_{r_{\rm in}}\int^{2\pi}_{0}
\frac{{\cal R}z(z-r~cot\theta_\s)^2rd{\phi}dr}{[(z-r~cot\theta_\s
)^2+r^2]^{5/2}\gamma^4_{\s}\left(1+{\rm v}_il^i\right)^4_\s}
\label{ppsd.eq}
\end{equation}
${\cal S}$ is a constant, which is 
obtained after converting the whole expression in geometric units mentioned after equation \ref{entacc.eq},
$$
{\cal S}=\frac{1.3{\times}10^{38}{\ell}_\s{\sigma}_{T}}{2{\pi}c m_e{A}_\s GM_{\odot}} 
$$
${\sigma}_{T},m_e,G,M_{\odot},{A}_\s,{\ell}_\s$ are Thomson 
scattering cross section, rest mass of electron, constant of 
gravitation, solar mass, surface area of PSD and post shock luminosity in units of 
Eddington luminosity ($L_{\rm Edd}=1.3\times 10^{38}M_B/M_\odot$) respectively.
Now, depending on the central mass, $\ell_\s$ is calculated from equation (\ref{chi8.eq} or \ref{chi1.eq}),
and hence depends on SKD and KD accretion rates.

The radiative moments from the SKD are
\begin{equation}
{\cal E}_{\sk}=\int^{r_{0}}_{r_{\rm l_1}}\int^{2\pi}_{0}F_{\sk}\frac{(r cos\theta_\sk+ d_0 sin\theta_\sk)zdrd\phi}
{ru^2_\sk\left(cot \theta_\sk r+d_0\right)^2[(z-r~cot\theta_\sk)^2+r^2]^{3/2}\gamma^4_{\sk}\left(1+{\rm v}_il^i
\right)^4_{\sk}}
\label{eskd.eq}
\end{equation}
\begin{equation}
{\cal F}_{\sk}=\int^{r_{0}}_{r_{\rm l_1}}\int^{2\pi}_{0}F_{\sk}\frac{(r cos\theta_\sk+ d_0 sin\theta_\sk)
z(z-r~cot\theta_{\sk})drd\phi}
{ru^2_\sk\left(cot \theta_\sk r+d_0\right)^2[(z-r~cot\theta_\sk)^2+r^2]^2\gamma^4_{\sk}\left
(1+{\rm v}_il^i\right)^4_{\sk}}
\label{fskd.eq}
\end{equation}
\begin{equation}
{\cal P}_{\sk}=\int^{r_0}_{r_{\rm l_1}}\int^{2\pi}_{0}F_{\sk}
\frac{(r cos\theta_\sk+ d_0 sin\theta_\sk)z(z-r~cot\theta_{\sk})^2drd\phi}
{ru^2_\sk\left(cot \theta_\sk r+d_0\right)^2[(z-r~cot\theta_\sk)^2+r^2]^{5/2}\gamma^4_{\sk}
\left(1+{\rm v}_il^i\right)^4_{\sk}}
\label{pskd.eq}
\end{equation}
With 
\begin{equation}
F_{\sk} = {\cal S}_{\rm ks}\left(\frac{u_0r_0H_0}{u_\sk rH}\right)^{
3\left(\Gamma-1\right)} \\ \nonumber
+{\cal S}_{\rm kb}\left(\frac{u_0r_0H_0}
{u_\sk rH}\right)^{\frac{\left(\Gamma-1\right)}{2}}
\left[1+1.78\left\{\Theta_0\left(\frac{u_0 r_0H_0}{u_\sk rH}\right)^{(\Gamma-1)}\right\}^{1.34}\right]
\label{facskd.eq}
\end{equation}
Constants ${\cal S}_{\rm ks}$ and ${\cal S}_{\rm kb}$ are 
associated with synchrotron and bremsstrahlung terms and are given as
$$
{\cal S}_{\rm ks}=\frac{9.22\times 10^{33}e^4\Theta_0^3\beta
\sigma_T{\dot m}^2_\sk}{\pi m^2_em_{p^+}^2c^2 G^2M^2_{\odot}}
$$

$$
{\cal S}_{\rm kb}=\frac{1.51\times 10^5 {\dot m}^2_\sk \sqrt{\Theta_0} \sigma_T}
{\pi^2 m^2_{p^+} M^2_{\odot}G^2 \sqrt{m_e k}}
$$
Here ${\dot m}_\sk={\dot M}_\sk/{\dot M}_{\rm Edd}$, where the Eddington accretion rate
${\dot M}_{\rm Edd}=1.44\times10^{17}(M_B/M_{\odot})$ g s$^{-1}$ and $L_{\rm Edd}={\dot M}_{\rm Edd}c^2$.
From a certain point $z$ on the axis, due to the shadow effect of the PSD the inner edge of SKD observed is given by
$$
r_{\rm l_1} (z)=\frac{z-d_0}{(z-H_s)/x_s+cot \theta_\sk}
$$
Therefore, to obtain the radiative moments at some $z$ on the
jet axis the integration limit on $r$ is $r_{\rm l_1}$ to $r_0$.

Similarly from KD, the moments are
\begin{equation}
{\cal E}_{\kd}={\cal K}\int^{r_0}_{r_{\rm l_2}}{\int}^{2{\pi}}_{0}
\frac{z(r^{-2}-{\sqrt {3}}r^{-5/2})d{\phi}dr
}{(z^2+r^2)^{3/2}
\gamma^4_{\kd}\left(1+{\rm v}_il^i\right)_\kd^4}
\label{ekd.eq}
\end{equation}
\begin{equation}
{\cal F}_{\kd}={\cal K}\int^{r_0}_{r_{\rm l_2}}{\int}^{2{\pi}}_{0}
\frac{z(r^{-2}-{\sqrt {3}}r^{-5/2})d{\phi}dr
}{(z^2+r^2)^{2}
\gamma^4_{\kd}\left(1+{\rm v}_il^i\right)_\kd^4}
\label{fkd.eq}
\end{equation}
\begin{equation}
{\cal P}_{\kd}={\cal K}\int^{r_0}_{r_{\rm l_2}}{\int}^{2{\pi}}_{0}
\frac{z(r^{-2}-{\sqrt {3}}r^{-5/2})d{\phi}dr}{(z^2+r^2)^{5/2}
\gamma^4_{\kd}\left(1+{\rm v}_il^i\right)_\kd^4}
\label{pkd.eq}
\end{equation}

The shadow effect of 
blocking a fraction of radiation from the KD by the PSD is also 
taken into account and the integration is done from $r_{\rm l_2}$
and is given by
$$
r_{\rm l_2}(z)=\frac{x_sz}{z-H_s},
$$
and the dimensionless constant ${\cal K}$ is given by
$$
{\cal K}=\frac{4.32{\times}10^{17}{\dot m_\kd}{\sigma}_Tc}{32
{\pi}^2m_eGM_{\odot}},
$$
with ${\dot m_\kd}$ being the Keplerian accretion rate in units of ${\dot M}_{\rm Edd}$.


\subsubsection{Nature of radiative moments}
\label{sbsbsec2.2.3}
\begin {figure}
\begin{center}
 \includegraphics[width=12.cm]{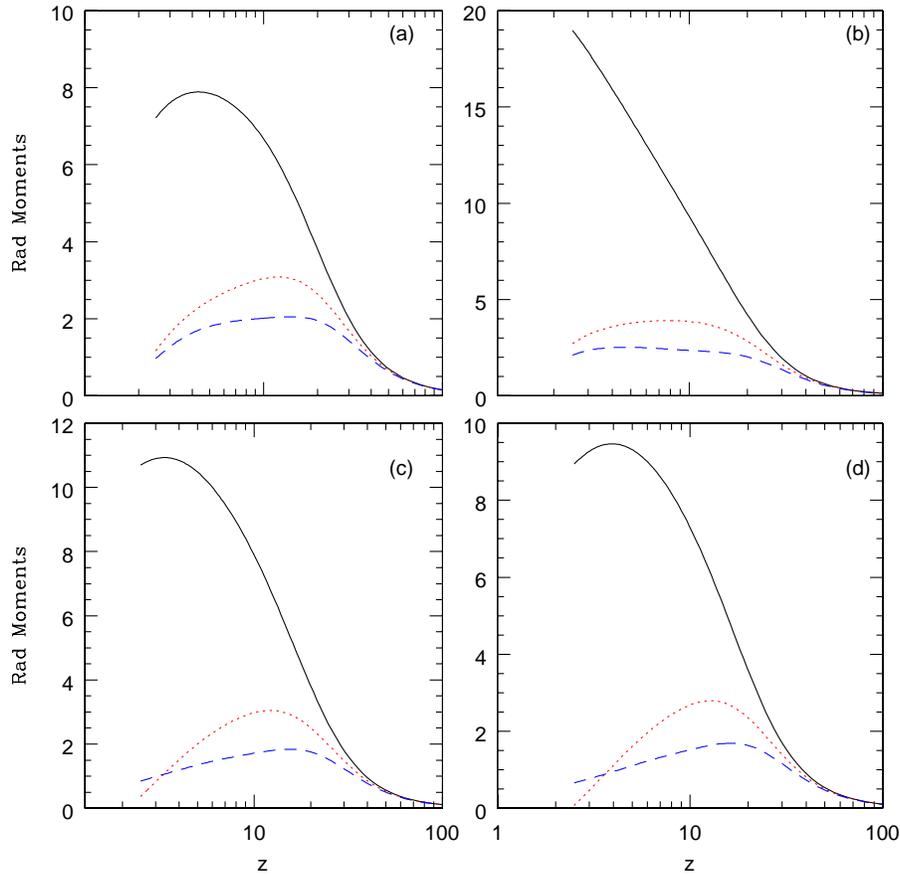}
 \caption{Distribution of radiative moments from the post-shock
 disc (PSD) with $z$
(in units of $r_g$). Each curve represents
${\cal E}_{\s}/{\cal S}$ (solid, black online), ${\cal F}_{\s}/{\cal S}$
 (dotted, red online)
and ${\cal P}_{\s}/{\cal S}$ (dashed, blue online), respectively. 
Various panels are for radiative moments (a) without relativistic transformations
in the accretion disc, (b) with special relativistic 
transformations up to first order
in $v$, (c) with full special relativistic transformations 
and (d) with approximate general
relativistic correction and full special
relativistic transformations. All the figures are obtained for ${\dot m}_\sk=5.06$, which
produces a shock at $x_s=20$ around a $10M_\odot$ BH.}
\label{lab:R1}
\end{center}
\end{figure}

\begin{figure}
\begin{center}
 \includegraphics[width=12.cm]{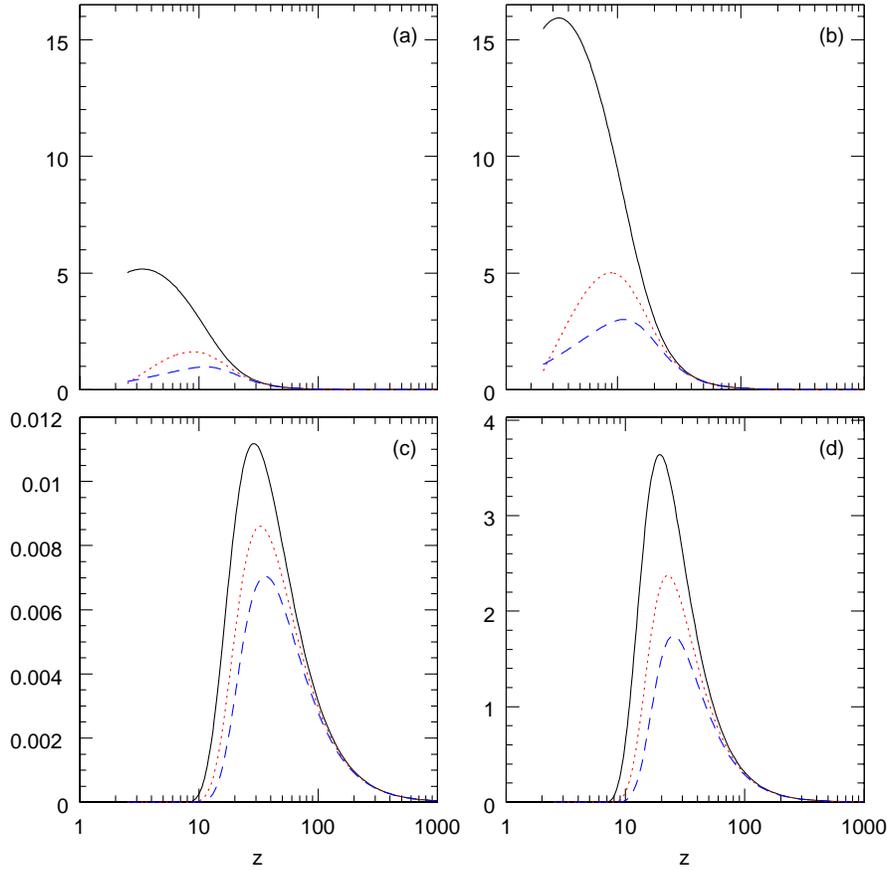}
\caption{Distribution of radiative moments  ${\cal E}$ (solid, online black), ${\cal F}$ (dotted, online red)
and ${\cal P}$ (dashed, online blue) from PSD for (a)
$10M_{\odot}$ and (b) $10^8M_{\odot}$ black holes.
Distribution of radiative moments from the
Kd (c) and from SKD (d) is same for both types
of black holes when expressed in the geometric units. Various parameters used to compute
the moments are ${\dot m}_\sk=7$, ${\dot m}_\kd=1$ and $\beta = 0.5$. This produce
$x_s= 13.2032$ and luminosities are ${\ell}_\sk=0.0265$, ${\ell}_\kd = 0.039$ and for stellar mass BH
${\ell}_\s = 0.215$ (a), 
while for larger BH $\ell_\s=0.661$
(b).}
\label{lab:R2}
\end{center}
\end{figure}

\begin{figure}
\begin{center}
 \includegraphics[width=12.cm]{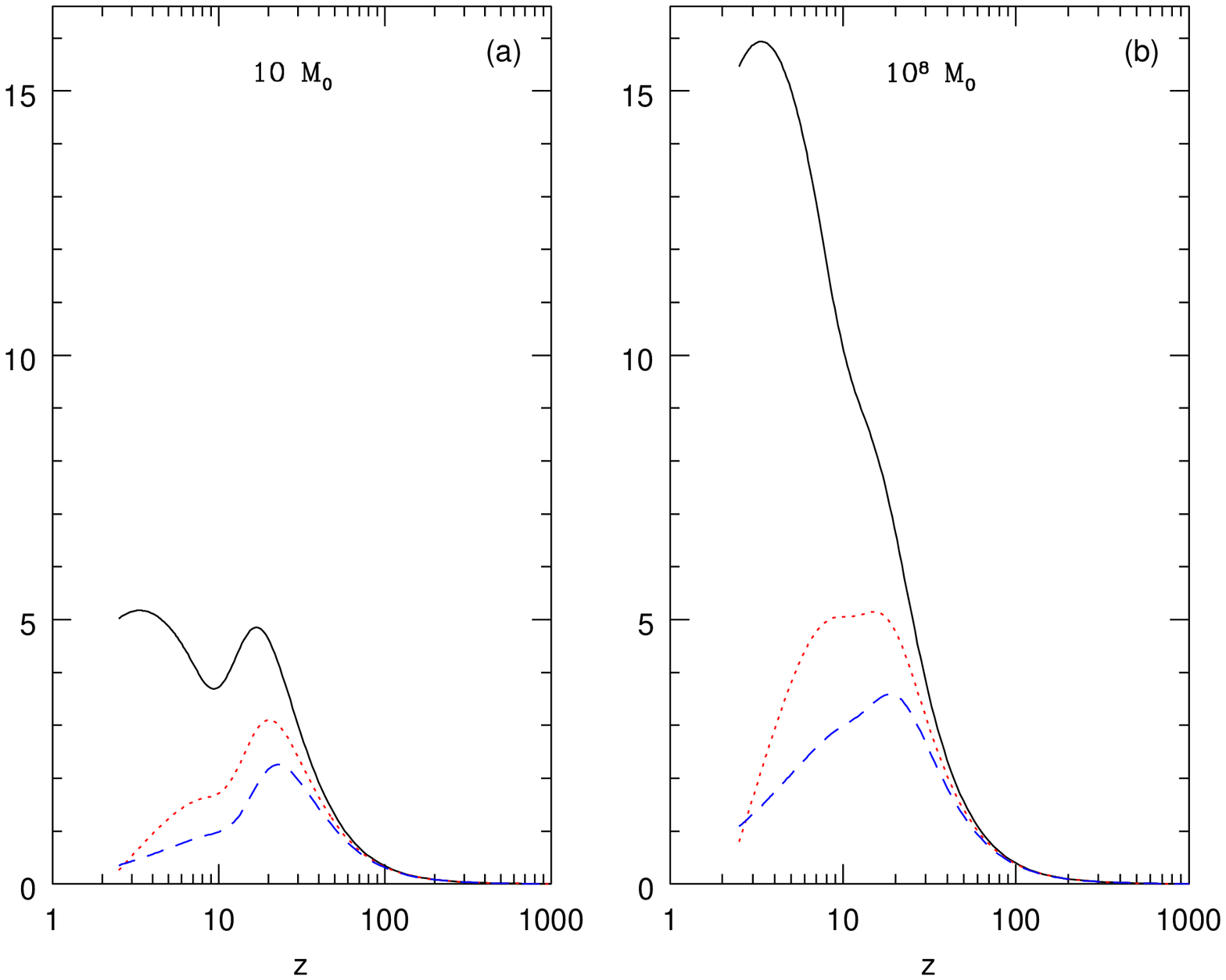}
 \caption{Combined moments from (a) $10$ and (b)
$10^8 M_{\odot}$ black holes. Various curves are ${\cal E}$ (solid, online black), ${\cal F}$
(dashed, online red) and ${\cal P}$ (dashed, online blue).
Disc parameters are ${\dot m}_\sk=7$ and ${\dot m}_\kd=1$, $\beta = 0.5$
for which the shock obtained is
at $x_s= 13.2032$, luminosities are $\ell_\sk=0.0265$,
$\ell_{\kd} = 0.039$. The $\ell_\s=0.215$ for $10M_\odot$ BH and $\ell_\s=0.661$ for $10^8M_\odot$ BH.
The moments are expressed in geometric units.}
\label{lab:R3}
\end{center}
\end{figure}

We now numerically integrate equations (\ref{epsd.eq}-\ref{pkd.eq}) to obtain the radiative
moments from PSD, SKD, and KD.
In Fig. \ref{lab:R1}a-d we present a comparative study of 
the space dependent part of the radiative moments {\ie}
${\cal E}_\s/{\cal S}$ (solid, black online), ${\cal F}_\s/{\cal S}$ (dotted, red online)
and ${\cal P}_\s/{\cal S}$ (dashed, blue online) from PSD which were computed in previous
works and the present one, because we would like to show the effect of various
corrections considered, while
computing the radiative moments. To compare the moments from
PSD, we need to know the size of PSD. This is obtained by estimating the $x_s$ from the
supplied ${\dot m}_\sk$. For ${\dot m}_\sk=5.06$, we estimate $x_s=20$ (equation \ref{xsdotm.eq}).
Figure \ref{lab:R1}(a) presents the radiative moments from PSD when 
the accretion disc was treated as a simple radiator without any dynamics
\citep{cdc04}. In Fig. \ref{lab:R1}(b), we applied special relativistic
transformations while computing the intensities in the observer frame from the local disc frame,
but the Lorentz factor 
($\gamma_{j}$) in the red-shift factor of the intensity (equation 
\ref{tran_int.eq}) was ignored \citep{c05}. And in Fig. \ref{lab:R1}(c)
we show moments computed with correct special relativistic transformations.
Figure \ref{lab:R1}(d) shows only the distribution of space dependent part of radiative moments from PSD
which is being used in the current study. In this panel, moments are with full special relativistic 
transformations and general relativistic correction for the radiation absorbed 
by the black hole near horizon. We notice that general 
relativistic correction (Fig. \ref{lab:R1}d) reduces the absolute value of the 
moments and the moments peak at a slightly different location on the axis
than the previous case (Fig.\ref{lab:R1}c). Ignoring the effect of disc motion while computing
radiative moments, under-estimates the moments (Fig. \ref{lab:R1}a), while
partial implementation of the very effect, over estimates the radiative moments (Fig. \ref{lab:R1}b) to unrealistic values.
In this paper, we use full special relativistic transformation and general relativistic corrections for the
radiative moments from PSD ( as in Fig. \ref{lab:R1}d).

From Appendix \ref{app:mdotxs2}, it is clear that $\ell_\s$ is different for $10M_{\odot}$ and $10^8M_\odot$
BH, for the same set of free parameters {\ie} $\dot m_\sk$ and $\dot m_\kd$. This should affect the net radiation field
above the disc. 
In Fig. \ref{lab:R2}a, we plot ${\cal E}_\s$ (solid, black online), ${\cal F}_\s$ (dotted, red online)
and ${\cal P}_\s$ (dashed, blue online) with $z$ for $\dot m_\sk=7$ and $\dot m_\kd=1$ for $M_B=10M_\odot$. 
For such accretion rate the shock is at $x_s=13.203$ (see, equation \ref{xsdotm.eq}). In Fig. \ref{lab:R2}b,
we plot radiative moments above a disc around a $M_B=10^8 M_\odot$ BH, for the same set of accretion parameters.
The radiative moments from the PSD around $10^8 M_\odot$ BH are about three times than
those around $10 M_\odot$. It may be noted, that for lower ${\dot m}_\sk$ the shocks are formed at larger
distance away from the BH, and the radiative moments around stellar mass and super-massive BH are same.
In Fig. \ref{lab:R2}c, we plot
${\cal E}_\kd$ (solid, black online), ${\cal F}_\kd$ (dotted, red online)
and ${\cal P}_\kd$ (dashed, blue online).
In Fig. \ref{lab:R2}d
we present ${\cal E}_\sk$ (solid, black` online), ${\cal F}_\sk$ (dotted, red online)
and ${\cal P}_\sk$ (dashed, blue online). The PSD luminosity for $10M_\odot$ BH is about $\ell_\s = 0.215$
and $\ell_\s=0.661$ for $10^8 M_\odot$ BH. The luminosities of the pre-shock disc $\ell_\sk=0.0265$ and
$\ell_\kd=0.039$ are same for discs around super massive, as well as, stellar mass BH.
The moments due to PSD around a super-massive BH is larger than that around stellar mass BH, however,
the SKD and KD contributions in the geometric units are exactly
same for stellar mass and super massive BH. In physical units these moments
would scale with the central mass.
Finally, we show combined
radiative moments from all the disc components for $10 M_\odot$ (Fig. \ref{lab:R3}a) and 
$10^8 M_\odot$ BH (Fig. \ref{lab:R3}b) for exactly the same disc parameters as in Fig. \ref{lab:R2}a-d. 
For higher ${\dot m}_\sk$, the overall radiation field (in geometric units), above a disc around a stellar mass 
BH is different than the moments around a super massive BH. This is because the for higher ${\dot m}_\sk$
the shock in accretion is located closer to the BH, which produces a cooler PSD around
a stellar mass BH than a super massive BH and therefore, larger efficiency of Comptonization.
However, for lower ${\dot m}_\sk$ (equation \ref{xsdotm.eq}
\& Fig. \ref{lab:figA1}a), the shock is located at larger distance from the BH, and the efficiency of
Comptonization is similar for both kinds of BH, and hence the moments are similar too.

\section{Solution method}
\label{sec3}
To obtain the solution of steady state, relativistic jet, we need to integrate equations (\ref{dthdr.eq}) 
and (\ref{dvdr.eq}) simultaneously. 
Alternatively, one may integrate equation (\ref{dvdr.eq}) with the help of equation (\ref{entacc.eq}),
since equation (\ref{entacc.eq}) is the integrated version of equation (\ref{dthdr.eq}).
We employ Runge-Kutta's $4^{th}$ order method
to integrate differential equations. 
\subsection{Sonic point conditions}
\label{sbsec3.1}
Since jets originate from a region in the accretion disc, close
to the central object, the base jet velocity should be small. However
due to this proximity of the jet base to the central object, the temperatures
at the jet base should be very high. In other words,
jets are subsonic at its base. While far away from the central object the thermal
energy and the radiative energy would drive jets to large $v$, but simultaneously becoming
less hot, {\ie} outer boundary condition of the jet is super sonic. This means at some point
the jet would become transonic, and the point in which this happens is called
the sonic point $z_c$ and the derivative $dv/dz\rightarrow 0/0$ (eq. \ref{dvdr.eq}).
This gives us the so-called sonic point conditions,
$$
 v_c=a_c
$$
\begin{eqnarray}
a^2_c=\frac{z_c}{2\gamma^2_c}\left[\frac{1}{2(z_c-1)^2}-\frac{\gamma^2_c\tau}{f_c+2\Theta_c}
\left\{(1+v^2_c){\cal F}_c-v_c({\cal E}_c+{\cal P}_c) \right\} \right].
\label{sonic.eq}
\end{eqnarray}

\begin{figure}
\begin{center}
 \includegraphics[width=12cm]{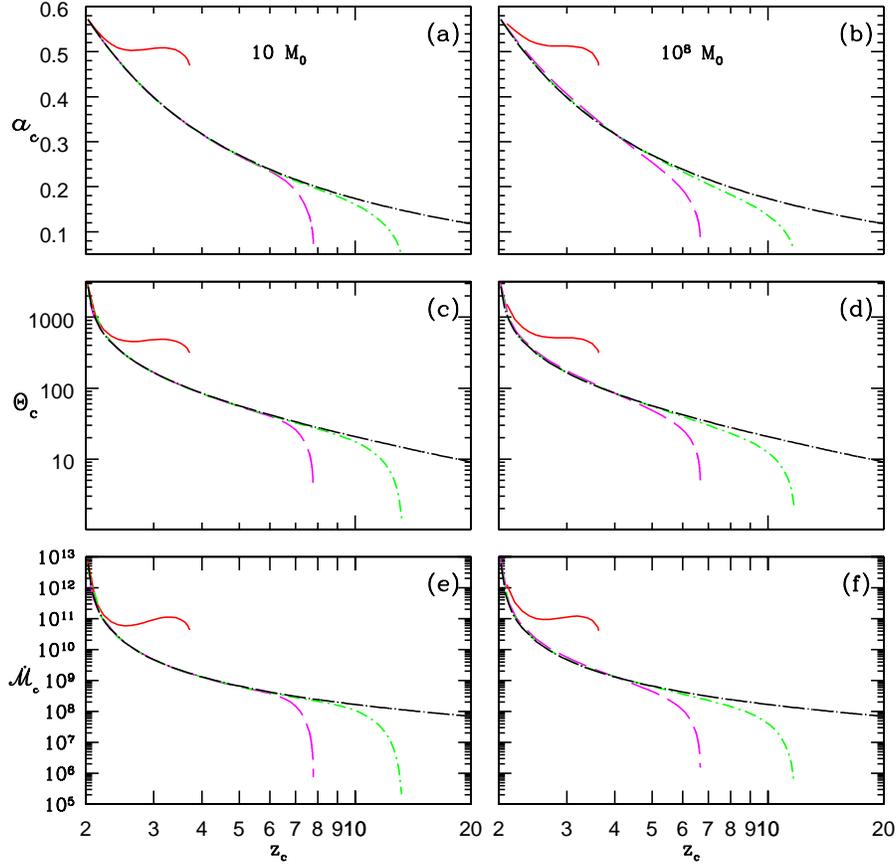}
 \caption{Variation of $a_c$ (a, b), $\Theta_c$ (c, d) and ${\dot {\cal M}}_c$ (e, f)
as a function of $z_c$ for jets are around $M1=10M_\odot$ BH (a, c, e) and $M8=10^8M_\odot$ BH (b, d, f).
Each curve corresponds
to ${\dot m}_\sk= 13$ (solid, red online), $10$ (long dashed, magenta online), $8$ (dashed-dotted, green online),
and only thermally driven jet (long-dashed dotted, black online). For all the plots ${\dot m}_\kd=1$.}
\label{lab:sonpt}
 \end{center}
\end{figure}

The $dv/dz|_c$ is calculated by employing the L'Hospital's rule at $z_c$ and solving the resulting
quadratic equation of $dv/dz|_c$. The resulting quadratic equation can admit two complex roots leading to
the so-called $O$ type or `centre' type
sonic points, or two real roots. The solutions with two real roots but with opposite signs
are called $X$ or `saddle' type sonic points, while real roots with same sign produces the nodal type sonic point. The
jet solutions flowing through X type sonic points are physical, and in this paper care has been taken to
study jet solutions through X type sonic points.
So for a given set of flow variables at the jet base, a unique solution will pass through
the sonic point determined by the entropy ${\dot {\cal M}}$ of the flow. For given
values of inner boundary condition {\ie} at the jet base  $z_b$, $v_b$ and $a_b$,
we integrate equation (\ref{dvdr.eq}) and (\ref{dthdr.eq}), while checking
for the sonic point conditions (equations \ref{sonic.eq}). We iterate till the
sonic point is obtained, and once it is obtained we continue to integrate outwards
starting from the sonic point. From equation (\ref{sonic.eq}) it is clear that for a thermally driven jet,
sonic point exists from $z_c=2~\rightarrow~ \infty$. However, radiatively driven flow may not posses
sonic point at large distances away from the jet base,
because the presence of strong radiation field may render $a_c\lsim 0$ at those distances.
In Figs. \ref{lab:sonpt}a-f, we compare the flow quantities $a_c$ (a, b), $\Theta_c$ (c, d), ${\dot {\cal M}}_c$
(e, f)
as a function of $z_c$. The left panels shows the sonic point properties of jets around $10M_\odot$ BH (a, c, e)
and the right panels show sonic point properties of jets around $10^8 M_\odot$ BH (b, d, f). The KD accretion rate, or,
${\dot m}_\kd=1$ is kept invariant for all these plots, but various curves are for ${\dot m}_\sk=13$
(solid, red online), $10$ (long dashed, magenta online), $8$ (dashed-dotted, red online),
and only thermally driven jet (long-dashed dotted, black online). It is interesting to note that,
the region outside the central object available for sonic points shrinks,
as the disc luminosity increases.
For luminous discs say ${\dot m}_\sk > 10$, sonic points can only form for $z_c < 8$. This implies that
only very hot flow has thermal energy density comparable to the radiation pressure, and therefore for any
flow with less thermal energy may be considered as collection of particles rather than a fluid in such radiation
field. Moreover, for ${\dot m}_\sk=13$, multiple sonic points may form for some values of
${\dot {\cal M}}$.

\section{Results}
\label{sec4}
\begin{figure}
\begin{center}
 \includegraphics[width=11cm]{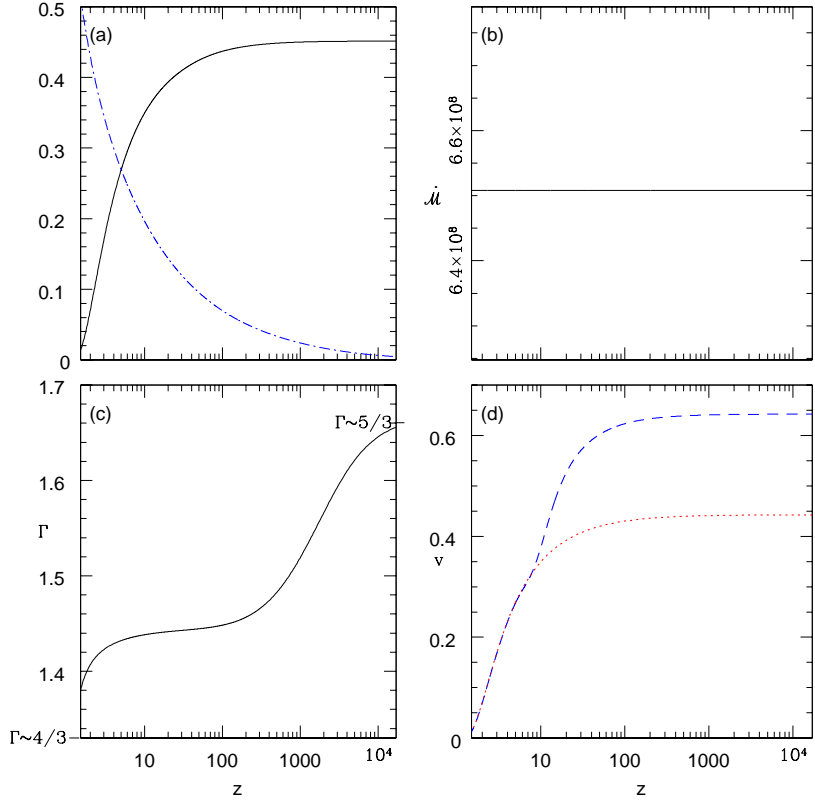}
 \caption{(a) The variation of jet 3-velocity $v$ (solid, black online) 
and sound speed $a$ (dashed-dotted, blue online) with $z$.
(b) ${\dot {\cal M}}$ is plotted as a function of $z$
and (c) the variation of $\Gamma$ with $z$. All the plots are generated
for inner boundary condition
$z_b=1.5$, $v_b=0.014$ and $a_b=0.51$, and the SKD and the KD accretion rates
 are ${\dot m}_\sk=5$
and ${\dot m}_\kd=1$, $\beta = 0.5$ and shock obtained is
at $x_s= 20.27 r_g$ and luminosities are ${\ell}_\sk=0.0082$, ${\ell}_\s = 0.113$,
 ${\ell}_{\kd} = 0.027$. (d) Comparison of $v$ for a thermally driven jet (dotted, red online) and
 radiatively plus thermally driven jet (dashed, blue online).
 Here accretion parameters are ${\dot m}_\kd=1$, $\beta = 0.5$ and ${\dot m}_\sk=10$ for the radiatively and thermally driven
 jet. For thermally driven flow no radiation interaction has been considered.
The luminosities are $\ell_\sk=0.108$, $\ell_\kd=0.055$ and $\ell_\s=0.38$.
 The composition of the jet is $\xi=1$ or $\ep$ plasma, and is launched around a $10M_\odot$ BH.}
\label{lab:gen}
 \end{center}
\end{figure}

A radiatively inefficient disc can only give rise to thermally driven jets and not radiatively driven
jets, so we choose luminous disc.
We discuss the jet properties for electron-proton jets {\ie} $\xi=1$, until specified otherwise.
We choose ${\dot m}_\kd=1$ to generate the KD radiative moments until specified otherwise.
The accreting material is assumed to posses stochastic magnetic field with constant
magnetic to gas pressure ratio $\beta=0.5$, until specified otherwise.
In Fig. \ref{lab:gen}a, we plot the jet 3-velocity $v$ (solid, black online)
and the sound speed $a$ (dashed dotted, blue online) as a function of $z$. This jet is from a disc
around a stellar mass BH.
The sonic point $z_c$ is at the crossing point of $v$ and $a$.
The terminal speed achieved for this case is $v_T\sim 0.45$, and the total disc
luminosity is $\ell=\ell_\s+\ell_\sk+\ell_\kd\sim 0.14$ in units of $L_{\rm Edd}$.
In Fig. \ref{lab:gen}b we plot ${\dot {\cal M}}$ as a function of $z$
and in Fig. \ref{lab:gen}c we plot the variation of $\Gamma$ or adiabatic index
of the jet. The inner boundary condition is $z_b=1.5 r_g$, $v_b=0.014$ and $a_b=0.51$,
and the SKD and the KD accretion rates are ${\dot m}_\sk=5$
and ${\dot m}_\kd=1$. Since the interaction between radiation and the jet material
is assumed to be in the Thompson scattering regime, the source term of
the first law of thermodynamics turns out to be zero ({\ie} equation \ref{dthdr.eq}),
and therefore ${\dot {\cal M}}$ which is a measure of entropy, remains constant
through out the flow. The base of the jet is very hot, therefore $\Gamma \rightarrow 4/3$
at the base. However, as the jet expands to relativistic velocities (at $z\rightarrow$ large), the temperature falls
such that $\Gamma \rightarrow 5/3$. In Fig. \ref{lab:gen}d, we compare the $v$ profile of a thermally driven jet (dotted, red online)
with a radiatively plus thermally driven jet (dashed, blue online) starting with the same base values. The radiatively
driven fluid jet (blue dashed) is powered by radiation from a disc with parameters ${\dot m}_\sk=10$
and ${\dot m}_\kd=1$. From the base to first few $r_g$, the $v$ profiles of the two flows are almost identical,
and the radiative driving is perceptible at $z>7.66$. The terminal speed of the thermally driven flow is slightly less than $0.45$
and for the radiatively driven flow it is $v_T\lsim 0.65$. The radiative driving of the jet is ineffective in regions
close to the $z_b$, because the thermal driving accelerates the jet to $v \sim v_{\rm eq}$. Therefore
radiative driving is ineffective in those region, and results in a similar $v$ profile up to $z\sim 7.66$,
but beyond it radiative driving generates a flow with a $44 \%$ increase in $v_T$.

\begin{figure}
\begin{center}
 \includegraphics[width=10.cm]{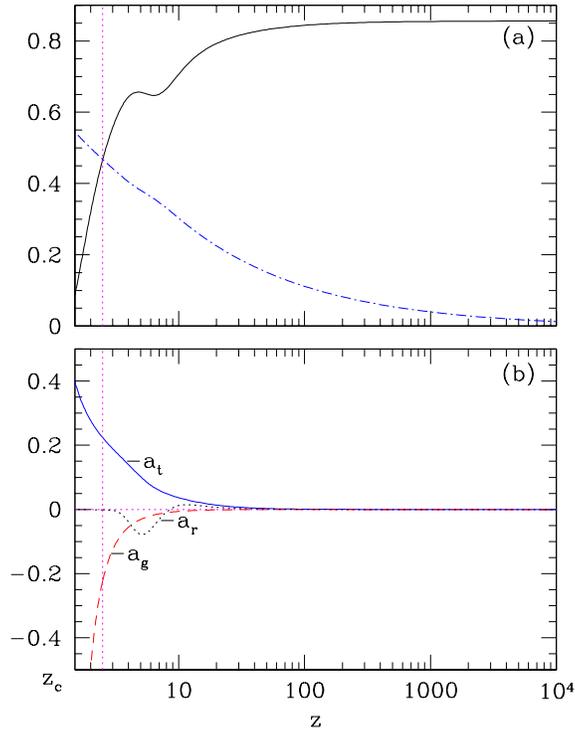}
 \caption{(a) Variation of $v$ (solid, online black) and $a$ (dashed-dotted, online blue),
The base speed $v_b=0.087$ and sound speed $a_b=0.545$ at jet base
 $z_b=1.5 r_g$, critical
 point obtained is at $z_c=2.5 r_g$; (b) Variation of ${\rm a}_t$ (solid, blue online),
${\rm a}_r$ (dotted, black online) and ${\rm a}_g$ (dashed, red online)
with $z$. The SKD and the
 KD accretion rates are ${\dot m}_\sk=12$
and ${\dot m}_\kd=1$, $\beta = 0.5$ and shock obtained is
at $x_s= 5.87$ and luminosities of various disc components around
$10M_\odot$ BH, are $\ell_\sk=0.295$, $\ell_{\s} = 0.522$ and
$\ell_{\kd} = 0.0667$. The composition of the jet is $\xi=1$. The vertical dashed line (magenta online)
shows the position of sonic point $z_c$.}
\label{lab:terms}
 \end{center}
\end{figure}

In Fig. \ref{lab:terms}a-b, we show a transonic jet from a disc around a stellar mass BH,
in which, the jet has been accelerated, as well as, decelerated by radiation.
In Fig. \ref{lab:terms}a, we plot $v$ (solid, black online)
and $a$ (dotted dashed, blue online) as a function of $z$ for inner boundary condition
$v_b=0.087$, $a_b=0.545$ at $z_b=1.5$, and ${\dot m}_\sk=12$ and ${\dot m}_\kd=1$.
In this case the sonic point is obtained at 
$z_c=2.5$. The total luminosity turns out to be $\ell=0.884$. 
The 3-velocity $v$ increases beyond $z_c$ and up to $z\sim 4$, and then decelerates
in the region $4< z \lsim 7$ and thereafter again accelerates till it reaches terminal value
at $v_T \sim 0.86$. Let us analyze the various terms that influence $v$.
In Fig. \ref{lab:terms}b, we plot the 
variation of gravitational acceleration term or a$_g$ (dashed, red online), the radiative term a$_r$ (dotted, black online),
and acceleration due to thermal
driving a$_t$ (solid, blue online). In both the panels the dashed vertical line (magenta online) shows the location of $z_c$.
From the l. h. s of equation (\ref{dvdr.eq}), it is clear that in the subsonic
region $v$ can increase ({\ie} jet accelerate)
with $z$ only if r. h. s is negative. While in the supersonic region the jet accelerates if
the r. h. s is positive. The gravity term or a$_g$ is always negative,
while a$_t$ is always positive. In this particular solution
a$_r<0$ for $z<8.53$.
In the sub sonic region {\ie} $z<z_c$,
$|{\rm a_r}| \ll {\rm a}_t$ and a$_t<{\rm a}_g$, therefore r. h. s of equation (\ref{dvdr.eq})
is negative and the jet is accelerated. At the sonic point a$_t={\rm a}_g+{\rm a}_r$.
For $z> z_c$, a$_r$ decreases to its minimum value at $z=5.15$. Gravity is less important at these
distances and a$_r\gsim {\rm a}_t$, which makes the r. h. s negative. Therefore, the jet decelerates in the range $4.78<z<6.34$.
For $z>6.34$, $|{\rm a}_r|$ decreases, ultimately becomes positive, making r.h.s to be positive again.
So the jet starts to accelerate at $z>6.34$ until $v\rightarrow v_T$.

\begin {figure}
\begin{center}
 \includegraphics[width=12.cm]{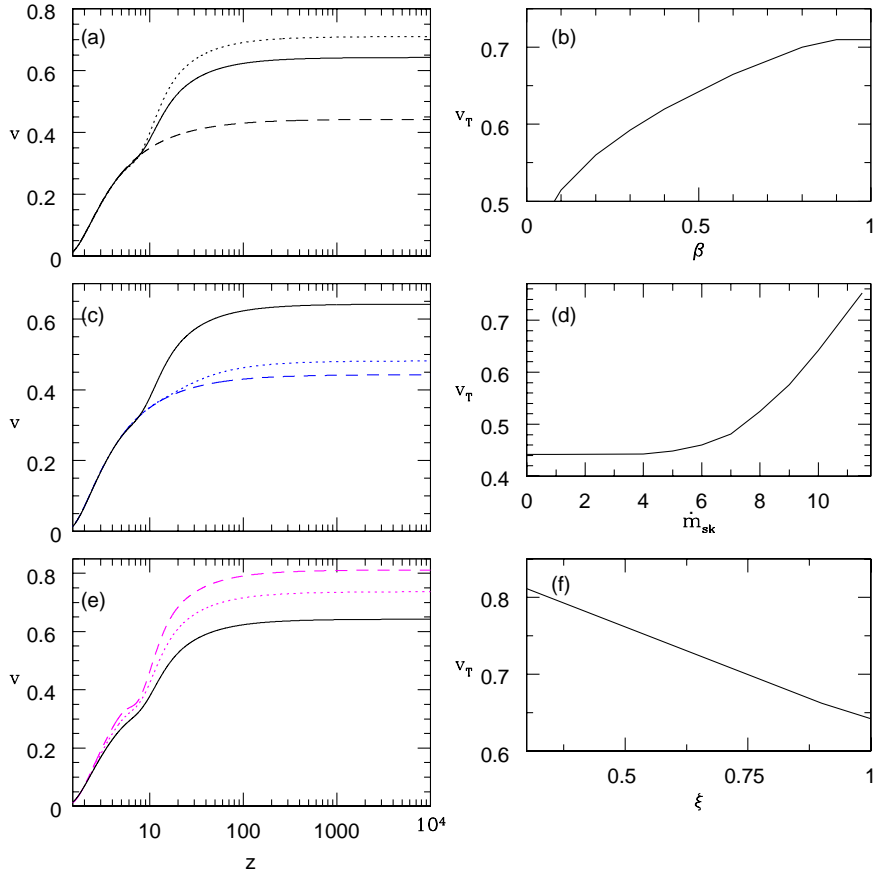}
 \caption{(a) Variation of $v$ with $z$. Each curve represents $\beta=1.0$ (dotted, black online),
$0.5$ (solid, black online) and
$0.0$ (dashed, black online).
(b) Variation of $v_{\rm T}$ with $\beta$. Other parameters same as (a).
(c) Variation of $v$ with $z$. Each curve represents ${\dot m}_\sk=1.0$ (dashed, blue online),
 $7$ (dotted, blue online) and
$10$ (solid, black online).
(d) Variation of $v_{\rm T}$ with ${\dot m}_\sk$. Other parameters same as (c).
(e) Variation of $v$ with $z$. Each curve represents $\xi=1.0$ (solid, black online),
$0.6$ (dotted, magenta online) and $0.3$ (dashed, magenta online).
(f) Variation of $v_{\rm T}$ with $\xi$. Other parameters same as (e).
Accretion parameters are ${\dot m}_\sk=10$ (in a, b, e, f), $\beta=0.5$
(in c, d, e, f), $\xi=1.0$ (in a, b, c, d) and ${\dot m}_\kd=1$. Jet base values
are $z_b=1.5$, $v_b=0.014$ and $a_b=0.51$.
The mass of central black hole is $10M_{\odot}$.}
\label{lab:parameters}
 \end{center}
\end{figure}

Now we discuss how various disc parameters and fluid composition 
of the jet affect dynamics of the jet. The jet is being affected by the
radiation from the disc, and the radiation field above
the disc in influenced by 
$\beta$, $\dot{m}_\sk$, and $\dot{m}_\kd$. 
The base values of the jet are $v_b=0.014$
and $a_b=0.51 $ at $z_b=1.5$.
In Fig. \ref{lab:parameters}a 
we show the comparison of $v$ as a function of $z$ for various values of ${\beta}
\rightarrow$ 0.0 (dashed, black online), 0.5 (solid, black online) and 1.0 (dotted, black online).
The corresponding terminal speeds
($v_T$) at $z=10^4$ with $\beta$ are presented in Fig. \ref{lab:parameters}b.
As the magnetic pressure increases in the disc {\ie} $\beta$ increases, the supersonic part of the jet
is accelerated because $\ell_\sk$ increases. When magnetic pressure 
is zero ($\beta=0.0$) {\ie} the jet is thermally and radiatively driven only by
pre-shock bremsstrahlung and thermal photons, the terminal speed is at around $0.44$. And when 
magnetic pressure is taken to be equal to the gas pressure, then $v_T$ increases above
$0.7$. 
In Fig. \ref{lab:parameters}c we show the effect of 
accretion rate of SKD on the $v$ profile and the corresponding terminal speeds
are presented in Fig. \ref{lab:parameters}d as a function of ${\dot m}_\sk$.
We see that $v_T$ ranges from 0.42 to 0.72 when the ${\dot m}_\sk$
is varied from 0.1 to 11.5. The velocity profile of a thermally driven jet, and a jet driven by radiations
acted on by ${\dot m}_\sk=1$ is similar. Only when the luminosity is close to $L_{\rm Edd}$ of $\ell \rightarrow 1$,
the radiative driving is significant. In Fig. \ref{lab:parameters}e we carry out 
similar analysis for the variation of composition parameter 
($\xi$) in the jet, and plot $v$ profiles for jet with $\xi=0.3$ (dashed, magenta online), $0.5$ (dotted, magenta online) and $1.0$
(solid, black online). With the lighter jet $v$ increases, and
this is also seen in the $v_T$ dependence of $\xi$ in
Fig. \ref{lab:parameters}f. As $\xi$ increases, high proton fraction makes the jet
heavier per unit pair of particles, and the optical depth decreases
due to the decrease in total number of leptons. So the net radiative momentum deposited on to the 
jet per unit volume decreases, in addition the inertia also increases.
This makes the jets with higher $\xi$ to be slower. 
\begin{figure}
\begin{center}
 \includegraphics[width=12.cm]{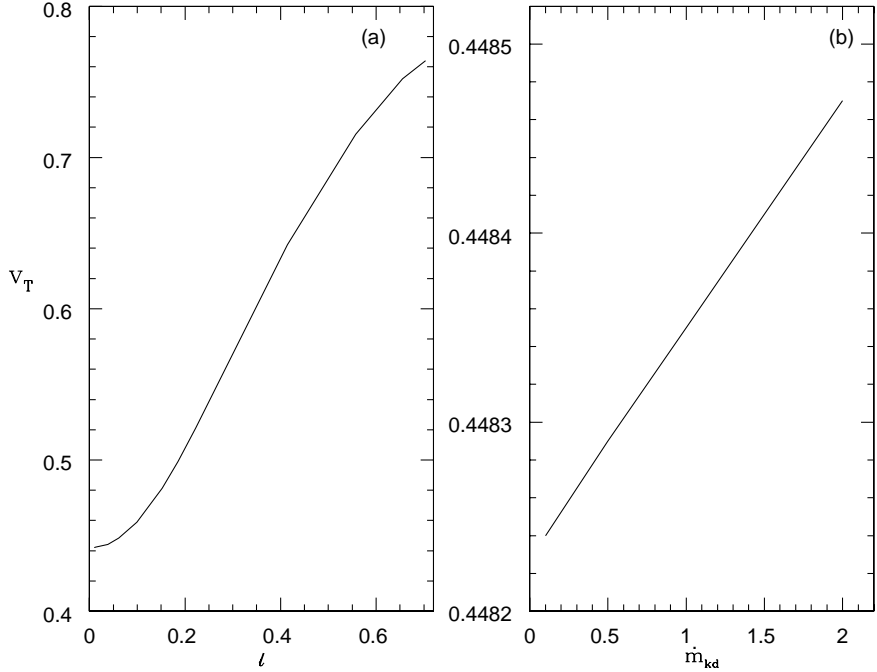}
 \caption{Dependence of terminal speeds $v_T$ on (a) $\ell$ and on
(b) $\dot{m}_{kd}$ for the same base values of jet ($z_b=1.5$, $v_b=0.014$. $a_b=0.51$).
The value of ${\dot m}_\kd=1$ for (a), and ${\dot m}_\sk=5$ for (b). While $\beta=0.5$ for both the cases.}
\label{lab:lumkd}
 \end{center}
\end{figure}

For various values of ${\dot m}_\sk$, $x_s$ changes and therefore not only $\ell_\sk$ changes but $\ell_\s$
changes too. Infact since $x_s$ is the inner edge of the KD, $\ell_\kd$ will change even though ${\dot m}_\kd$
is kept constant.
In Fig. \ref{lab:lumkd}a, we plot $v_T$ with the total luminosity $\ell~(\equiv \ell_\s+\ell_\sk+\ell_\kd)$,
by tuning ${\dot m}_\sk$. As the luminosity of the disc increases the terminal speed increases from moderate
values of $0.44$ to high speeds of $\sim 0.8$ when the disc luminosity is closer to Eddington limit.
However, $\ell_\kd$ has limited role in determining $v_T$ as has been shown in Fig. \ref{lab:lumkd}b.

\begin {figure}
\begin{center}
 \includegraphics[width=14.cm]{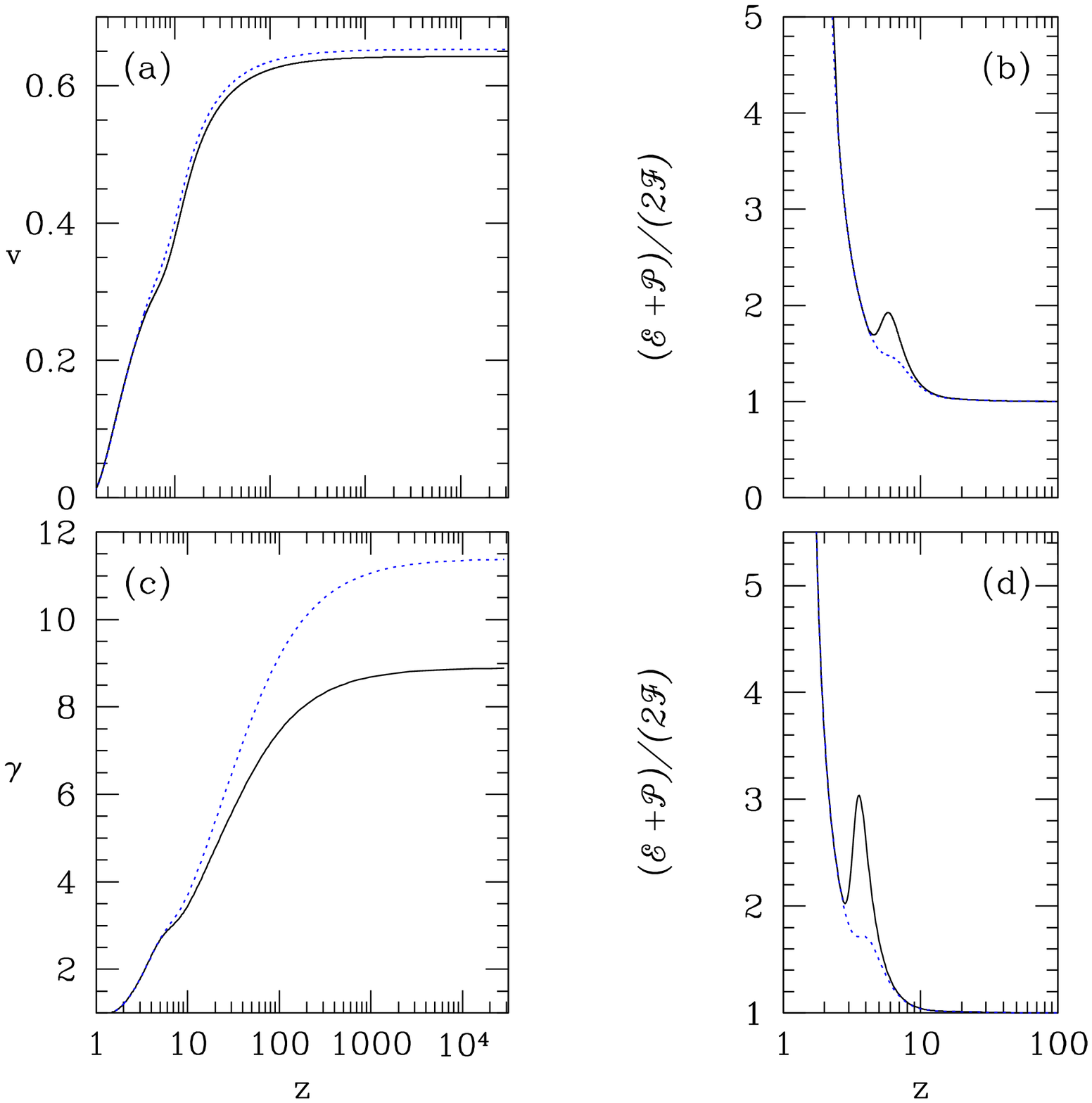}
 \vskip 0.0cm
 \caption{ Comparison of (a) $v$ profile for jets around $10 M_\odot$ (solid, black online)
and $10^8 M_\odot$ black holes (dotted, blue online).
The SKD accretion rate is ${\dot m}_\sk=10$.
(b) $\Re=({\cal E}+{\cal P})/2{\cal F}$ with $z$
from an accretion disc around $10 M_\odot$ BH (solid, black online)
and $10^8 M_\odot$ BH (dotted, blue online). 
The jet base values for (a) and (b) are $z_b=1.5$, $v_b=0.014$, and $a_b=0.51$.
(c) $\gamma$ profile of
a jet around $10M_\odot$ BH (solid, black online) and $10^8M_\odot$ (dotted, blue online).
(d) $\Re=({\cal E}+{\cal P})/2{\cal F}$ with $z$.
The SKD parameter is ${\dot m}_\sk=12$. Other disc parameters are
${\dot m}_\kd=1$, $\beta=0.5$. And the jet base values for (c) and (d) are $z_b=1.5$, $v_b=0.19$, and $a_b=0.576$.}
\label{lab:m1m8}
 \end{center}
\end{figure}

In Fig. \ref{lab:R3} the radiative moment around a super-massive BH is shown to be significantly higher than
that around a stellar mass BH even if the accretion rates (in units of ${\dot M}_{\rm Edd}$) are same.
In order to study the effect of the mass of the central object, in 
Fig. \ref{lab:m1m8}a, we compare the $v$ profile of the jet around $10M_\odot$ BH (solid, black online)
with that around $10^8M_\odot$ BH (dotted, blue online). The jets are launched with the same
base values ($z_b=1.5$, $v_b=0.014$, and $a_b=0.51$), and the accretion rates in units of ${\dot M}_{\rm Edd}$
are exactly same. Although the radiative moments around a super-massive BH are significantly different,
yet the $v$ profiles differ by moderate amount. To ascertain the cause we plot $\Re$ or relative contribution of
radiative moments for both the jets in Fig. \ref{lab:m1m8}b. $\Re$ is quite similar
for both the BHs close to the horizon, but in the range $4<z<10$ $\Re$ around stellar mass BH is higher than that around
super massive BH. In Fig. \ref{lab:m1m8}c, we compare the Lorentz factor
$\gamma$ of a jet around $10M_\odot$ BH (solid, black online) with a jet $10^8M_\odot$ (dotted, blue online),
launched with hot base ($z_b=1.5,~v_b=0.19,~
a_b=0.576$) and acted by high accretion rate ${\dot m}_\sk=12$. The initial jet $\gamma$ ($\equiv$ $v$) is almost
same for both the jets, however, due to larger $\Re$ around a stellar mass BH, the jet around it slower
compared to that around super-massive BH. In this case, the terminal Lorentz factor $\gamma_T$
is significantly larger for a jet around $10^8 M_\odot$ BH.

\begin {figure}
\begin{center}
 \includegraphics[width=12.cm]{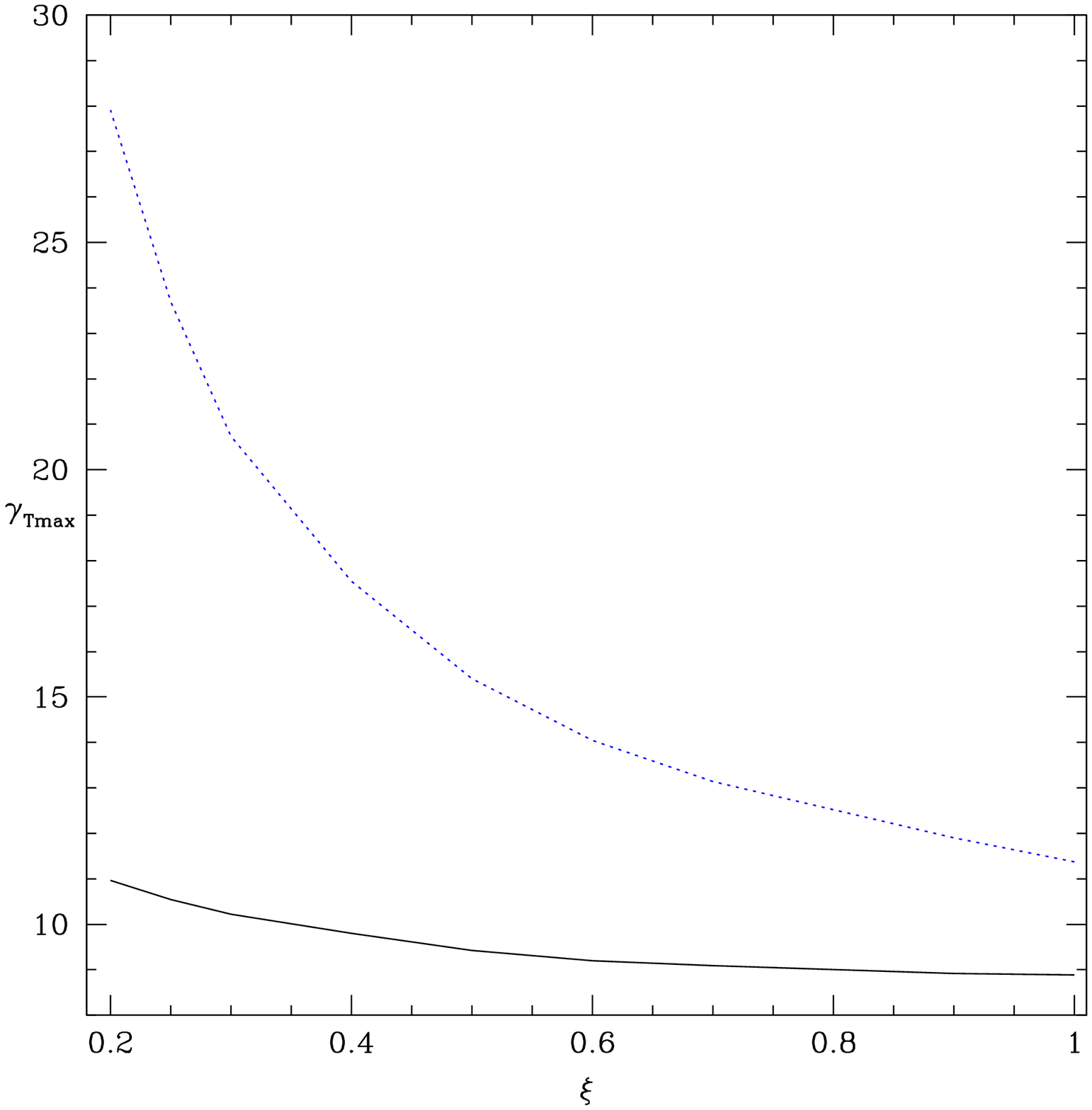}
 \caption{Variation of $\gamma_{T{\rm max}}$ with $\xi$, around $10M_\odot$ BH (solid, black online) and
$10^8 M_\odot$ BH (dotted, blue online). Accretion parameters are ${\dot m}_\sk=12$, $\beta=0.5$ and ${\dot m}_\kd=1$.
and jet base values are $z_b=1.5,~\&~a_b=0.5766$.}
\label{lab:max1}
 \end{center}
\end{figure}

It is clear that jets around stellar mass BH are slower, and lighter jets are faster, but what is the maximum
terminal velocity possible? We choose to launch jet with maximum possible sound speed at the base and very high
accretion rate.
In Fig. \ref{lab:max1},
we plot the maximum terminal Lorentz factor or $\gamma_{T{\rm max}}$ possible as a function of $\xi$ for $10M_\odot$ BH (solid, black online)
and $10^8M_\odot$ (dotted, blue online), when accretion parameters are ${\dot m}_\sk=12$, $\beta=0.5$ and ${\dot m}_\kd=1$.
For jet composition $\xi=1.0$, the maximum possible terminal Lorentz factor $\gamma_{T{\rm max}}\lsim 9$
for a jet around $10M_\odot$ BH, but for $10^8 M_\odot$ BH, $\gamma_{T{\rm max}}\gsim 11$. However, for lighter jet
around stellar mass BH $\gamma_{T{\rm max}}\sim 10$, but for super-massive BH, light jets yields
$\gamma_{T{\rm max}}\sim$ few $\times 10$. So for light jets ultra-relativistic jets around super-massive BH
is possible if it is driven by the radiation from a luminous disc.

\section{Discussion and Concluding Remarks}
\label{sec5}

In this paper, we have investigated the interaction of a relativistic fluid-jet with the radiation
field of the underlying accretion disc. The accretion disc plays an auxiliary role, in other words,
the jet-disc connection has been ignored in the present study. In principle the accretion rates,
and the outer boundary condition, determines the jet states, as well as, the radiation field around it.
We reduced the number of free parameters to determine the radiation field, by estimating the temperature,
density and the velocity profile of the various components of the accretion disc and considered them
as generic (Appendix A). The radiative
intensities of the SKD is determined from the estimated temperature,
density and the velocity profile by supplying the accretion rates of ${\dot m}_\sk$ (Appendix \ref{app:accvT}).
The same for KD is obtained only from the ${\dot m}_\kd$.
We used some known accretion disc solutions from the literature \citep{kc14}, to obtain a simplifying analytic relation
between the shock location and the SKD accretion rate (Appendix B). This enables us to reduce one free parameter ({\ie} $x_s$).
Now we use those same accretion disc solutions in the general radiative transfer code \citep{mc08}, and compute the
spectra, as well as, ratio of the luminosities from the PSD and the pre-shock disc as a function of shock location
(Appendix B).
In other words, the accretion rates determine the accretion solutions and the shock location, which determines
the disc luminosities, intensities and therefore the entire radiation field around it. And then we studied the propagation of
a transonic and relativistic jet through this radiation field. 

We noticed that proper relativistic transformation of the radiative intensities from the local disc frame
to the observer frame is very important and these transformations
modify the magnitude, as well as, the distribution of the
moments around a compact object. The PSD and SKD are the major contributors in
the net radiative moments and KD contribution is much lower than either one of the former two.
However, the contribution of PSD compared to that of SKD in the net moment, will depend on where the shock forms
which is also dictated by the accretion rate. One of the interesting fact about the moments due to various
disc components is that they peak at different positions away from the disc plane. Therefore the jet which
is initially thermally driven, but is further accelerated by the radiative moments from various disc components
further down stream.
The jets with normal conditions at the base, produces
mildly relativistic jets with terminal velocities $v_T\sim$ few $\times 0.1$ (Fig. \ref{lab:gen}).
The scattering regime maintains the isentropic nature of the jet, and because we considered a realistic
and relativistic gas equation of state, the adiabatic index changes along the jet.
However, radiation not only accelerates but also decelerates if $v>v_{\rm eq}$. Although close to the jet base
($z_b$) the velocity is low and the radiation field should accelerate, but being hot the effect of radiation
is not significant in the subsonic branch because of the presence of inverse of enthalpy term
in the radiative term ${\rm a}_r$ of the equation of motion (equation \ref{dvdr.eq}).
Therefore, in the subsonic domain the jet is accelerated as a result of competition between thermal and the gravity term.
In the supersonic domain the gravity term is smaller than both the thermal and the radiative term. 
If ${\rm a}_r<0$ and $|{\rm a}_r|\gsim {\rm a}_t$, then the jet can be decelerated (Figs. \ref{lab:terms}a,b). 
As the magnetic pressure is increased, the synchrotron radiation from SKD increases and it jacks up the
flow velocity in the supersonic regime. Increasing the ${\dot m}_\sk$ increases both the 
synchrotron, as well as, bremsstrahlung photons from the SKD, which makes the SKD contribution
to the net radiative moment more dominant, and therefore increases the $v$ in the supersonic
part of the flow. The $v_T$ increases with $\beta$ but tends to taper off as $\beta\rightarrow 1$,
however, $v_T$ tends to increase with ${\dot m}_\sk$ and shows no tendency to taper off.
The jet tends to get faster with the decrease in protons ({\ie} decrease of $\xi$).
We do not extend our study to electron and positron or $\xi=0$ jet, since a purely
electron-positron jet is highly unlikely \citep{kscc13}, although pair dominated
jet ({\ie} $0<\xi<1$) is definitely possible. So we extend our study to $0.27\leq \xi \leq 1$
jets (Fig. \ref{lab:parameters}a-f). Although the terminal velocity $v_T$ increases with the total
luminosity $\ell$, and approaches relativistic values as the disc luminosity approaches
the Eddington limit. However, the KD plays a limited role in accelerating jets (Figs. \ref{lab:lumkd} a,b).
Since the radiation field around a super-massive BH and stellar mass BH is different, we
compared jets starting with the same base values, same accretion rates (in units of ${\dot M}_{\rm Edd}$)
but around stellar mass and super-massive BH.
If accretion rates are moderately high, then jets around super-massive BHs are slightly faster than
those around stellar mass BH (Figs. \ref{lab:m1m8}a, b). However, if accretion accretion rates are
high such that the moments from the central region of the disc differ significantly, then ultra-relativistic
jets around super massive BHs can be obtained compared to stellar mass BH (Figs. \ref{lab:m1m8}c, d).
For high accretion rates, it has been shown that around stellar mass BH,
moments due to PSD and the pre-shock disc
are comparable, which makes $\Re > 1$ for a larger distance away from the BH (Fig. \ref{lab:R3}). This limits
the maximum velocity a jet may achieve around a stellar mass BH, compared to that around a super-massive BH.
This brings us to the issue of maximum possible terminal velocity of the jet. We launched jets with relativistic
sound speed in order to get maximum thermal driving, while increase the accretion rates to very high values.
The comparison of jet $\gamma_{T{\rm max}}$ around a stellar mass BH and that around super massive one
as a function of $\xi$,
shows that jet around super massive BH can be accelerated to $\gamma_T \gsim 11$ even for fluid composition
$\ep$, while that around stellar mass BH $\gamma_T \sim$ few. However, lighter jets around $10^8 M_\odot$ BH
can be accelerated to
truly ultra-relativistic speeds, compared to jets around stellar mass BH.

It is interesting that the radiative moments from different disc components maximize at different distances
from the jet base, which opens up the possibility
of multi stage acceleration of the jet. However,
close to the base, thermal acceleration is the main driving force that
makes the flow supersonic, and thereby negating the effect of gravity. Thereafter, the radiative driving accelerates jet
to mildly relativistic terminal velocities if the radiation field is mild, but intense radiation field
can accelerate jets to ultra-relativistic velocities. Hence, the thermal driving of the jet
at the base, along with the radiative driving further out, comprises a complete multi-stage acceleration
rather than just the radiation field itself.
Although the maximum terminal speeds achieved is truly ultra-relativistic, especially around
super-massive BHs, but at the jet base, which is hot, the disc photons may take away energy from the hot electrons
instead of imparting its momentum onto it. The jet base may also radiate via other processes.
These radiations would, in the observer frame, actually be flowing along the jet and therefore may interact
and deposit momentum on to the jet
further down stream. The jet may also gain energy via free-free absorption. These aspects have been
ignored and therefore the conclusions might slightly differ. However, it is clear the jet can be
accelerated to relativistic terminal speed.
We would like to simulate the present work of radiation driven relativistic jets around compact objects,
similar to our effort with galactic outflows \citep{csnr12}. One of the major difference expected is,
if the radiative states of the accretion disc changes, how does it affect the jet and what are the timescales
in which these changes are expected to be observed in the jets. These advances of our work are underway and
will be reported elsewhere.

\section*{Acknowledgment}%
The authors acknowledge the anonymous referee for helpful suggestions.

{}

\appendix
\section{Estimation of temperature and velocity of SKD and PSD}
\label{app:accvT}
\begin {figure}
\begin{center}
 \includegraphics[width=10.cm]{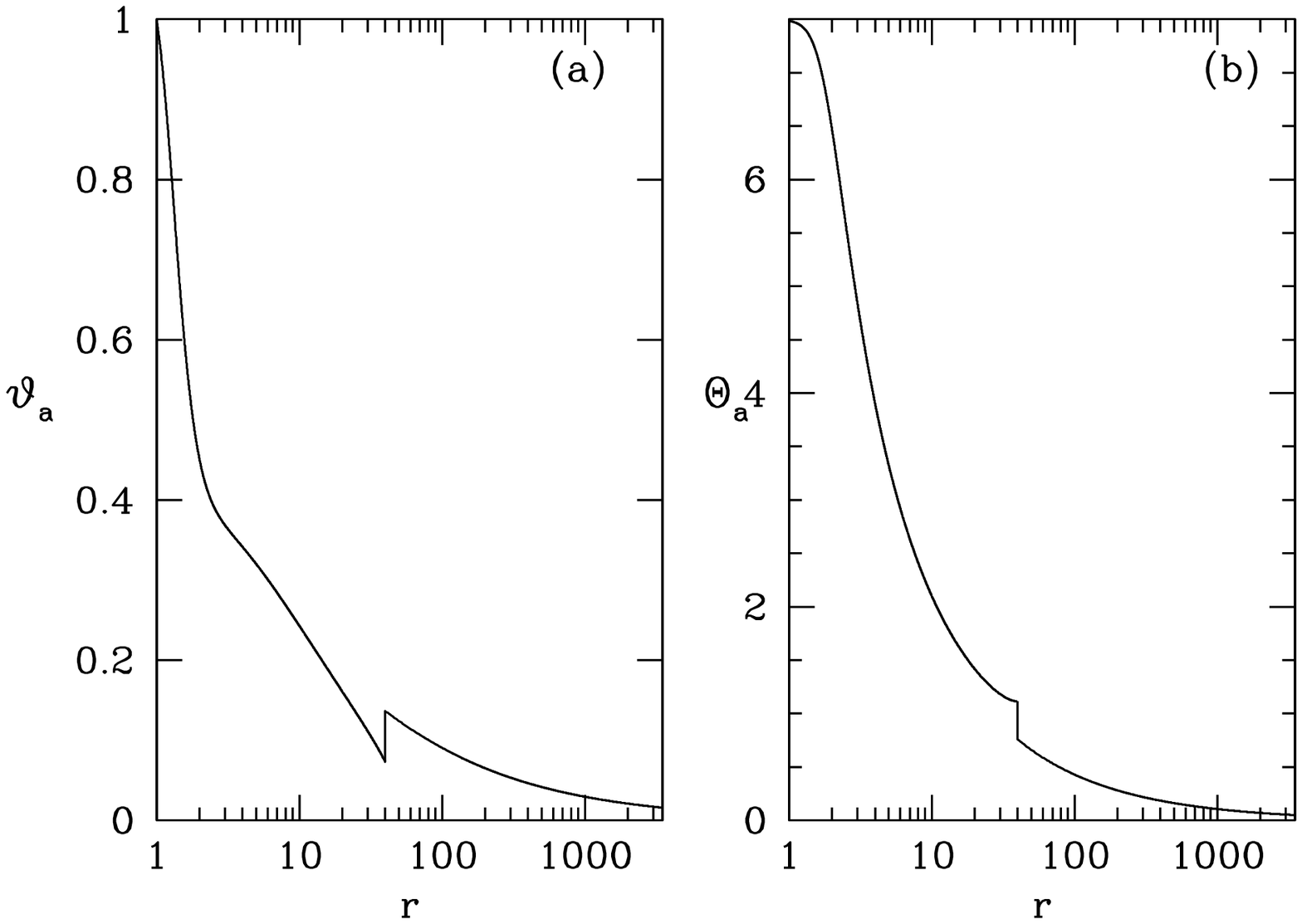}
 \caption{Velocity $\vartheta_{\rm a}$ and dimensionless temperature $\Theta_{\rm a}$ estimated for
accreting flow for angular momentum $\lambda=1.7$. Here $r_0=3500$ and $\vartheta_0=0.01$.}
\label{lab:figA1}
 \end{center}
\end{figure}

In the present analysis, the focus rests upon outflows and the 
accretion disc plays an auxiliary role, as radiation from the 
various accretion disc components affect the jet dynamics.
Since the accretion disc emission depends on the flow variables
of the disc, we need to know the temperature and velocity distribution.
The method to estimate the velocity distribution of SKD and PSD are given below.
We estimate the velocity from geodesic equations, which shows that the covariant
time component of four velocity {\ie} $u_t$ is a constant of motion.
From $u_\mu u^\mu=-1$, find
Now,
$$
-u_tu^t=-g^{tt}(u_t)^2=\gamma^2=\gamma_{\vartheta}^2 \gamma_\phi^2{\equiv}\mbox{square of Lorentz factor},
$$
where,
\be
 \gamma^2_{\vartheta_{\rm a}} = \frac{1}{1-\vartheta^2_{\rm a}};~  \gamma^2_\phi = \frac{1}{1-{\rm v}_{\phi \rm a}^2};
~{\rm v}_{\phi \rm a}^2=\frac{-u_\phi u^\phi}{u_t u^t}=
\frac{\left(r-1\right)\lambda^2}{r^3}.
\label{lofac.eq}
\ee
Here, $\vartheta_a$ is the radial 3-velocity measured by a co-rotating observer, while the radial 3-velocity is $v_{r \rm a}$,
and are defined as
$$
{\rm v}^2_{r{\rm a}}=\frac{-u_r u^r}{u_t u^t};~\&~\vartheta^2_{\rm a}=\gamma^2_\phi {\rm v}^2_{r{\rm a}}
$$
The suffix `a' stands for either PSD, SKD or KD. For KD ${\rm v}_{r{\kd}}=\vartheta_{\kd}=0$ and
${\rm v}_{\phi \kd}$ is the Keplerian azimuthal velocity.
We chose outer boundary conditions ($\vartheta_{\rm a}=\vartheta_{0\rm a}$, and $\lambda_0$ at $r=r_0$), 
this relation allows us to calculate $u_t$ for SKD,
\be
(u_t)^2|_{r_0}=\left(1-\frac{1}{r_0}\right)\frac{1}{1-\vartheta^2_{0 \rm a}}.\frac{r^3_0}
{r^3_0-(r_0-1)\lambda^2_0}.
\label{utacc.eq}
\ee
And then from equation \ref{utacc.eq} we obtain $\vartheta(r)_\sk$,
\be
\vartheta_\sk=\left[1-\frac{(r-1)r^2}{\{r^3-[(r-1)\lambda^2]\}u_t^2|_{r_0}}
\right]^{1/2}.
\label{velsk.eq}
\ee
Here, the equations are expressed in geometric units $2G=M_B=c=1$. One must note from equation (\ref{velsk.eq})
that as $r\rightarrow 1$, $\vartheta\rightarrow 1$, although SKD in presence of shock, does not extend upto
the horizon. However, one should also remember
since the velocity is obtained from geodesic equations, it is slightly over estimated because
the pressure gradient terms were ignored while estimating the velocity. This would under estimate
the radiative moments slightly and hence our results are believable since
there is no over estimation of jet and radiation interaction.
To get the velocity distribution for the PSD,
we assume the shock compression ratio is $3$,
so the ratio of post shock (suffix $+$) and preshock (with suffix $-$) velocities
are $\vartheta_+=\vartheta_-/3$ at $r=x_s$, with this boundary condition we recalculate the constant
of motion and then use it to estimate the $\vartheta$ distribution for PSD,
\be
\vartheta_\s=\left[1-\frac{(r-1)r^2}{\{r^3-[(r-1)\lambda^2]\}u_t^2|_{x_s}}
\right]^{1/2}.
\label{velpsd.eq}
\ee
In Fig. (\ref{lab:figA1}a) we have plotted the velocity distribution of SKD and PSD, where the suffix
`a' signifies either $\sk$ or $\s$.

\subsection{Density and temperature of SKD}
\label{app:accvT2}

The SKD accretion rate equation is
\be
{\dot M}_\sk \approx \rho_\sk u^r_\sk r H_\sk,
\label{acc.eq}
\ee
here $H_\sk$ is the local height of SKD.
If the outer boundary of the accretion disc is at $r_0$, the height at outer boundary $H_0$,
the radial four velocity $u^r_0$, the dimensionless
temperature $\Theta_0$
and the density $\rho_0$, then we have
\begin{equation}
 \frac{\rho_\sk}{\rho_0}=\frac{u^r_0r_0H_0}{u^r_\sk rH_\sk}
\label{densrat.eq}
\end{equation}
Here, $u^r_\sk$ is obtained
from equation \ref{velsk.eq} and $H_\sk=r cot\theta_\sk+d_0$.

Assuming fixed $\Gamma$ and moderate radiative loss, then the dimensionless temperature
or $\Theta_\sk=p_\sk/\rho_\sk c^2$ is given by,
\begin{equation}
 \Theta_\sk=\Theta_0\left(\frac{\rho_\sk}{\rho_0}\right)^{\Gamma-1}=\Theta_0\left(\frac{u^r_0r_0H_0}{u^r_\sk rH}
 \right)^{\Gamma -1}
\label{thetacc.eq}
\end{equation}
Equations (\ref{thetacc.eq} and \ref{densrat.eq}) are used to compute the intrinsic synchrotron and bremsstrahlung
intensity of the SKD. Using the expression of $u^r_{\rm a}=(1-1/r)^{1/2}\gamma_{\vartheta_{\rm a}}\vartheta_{\rm a}$,
in equation \ref{thetacc.eq}, the SKD and PSD temperature distribution can be obtained. In Fig. (\ref{lab:figA1}b),
the dimensionless temperature distribution $\Theta_{\rm a}$ is plotted for both SKD and PSD.

\section{Relation between $\dot m_\sk$ and $x_s$}
\label{app:mdotxs}
\begin {figure}
\begin{center}
 \includegraphics[width=12.cm]{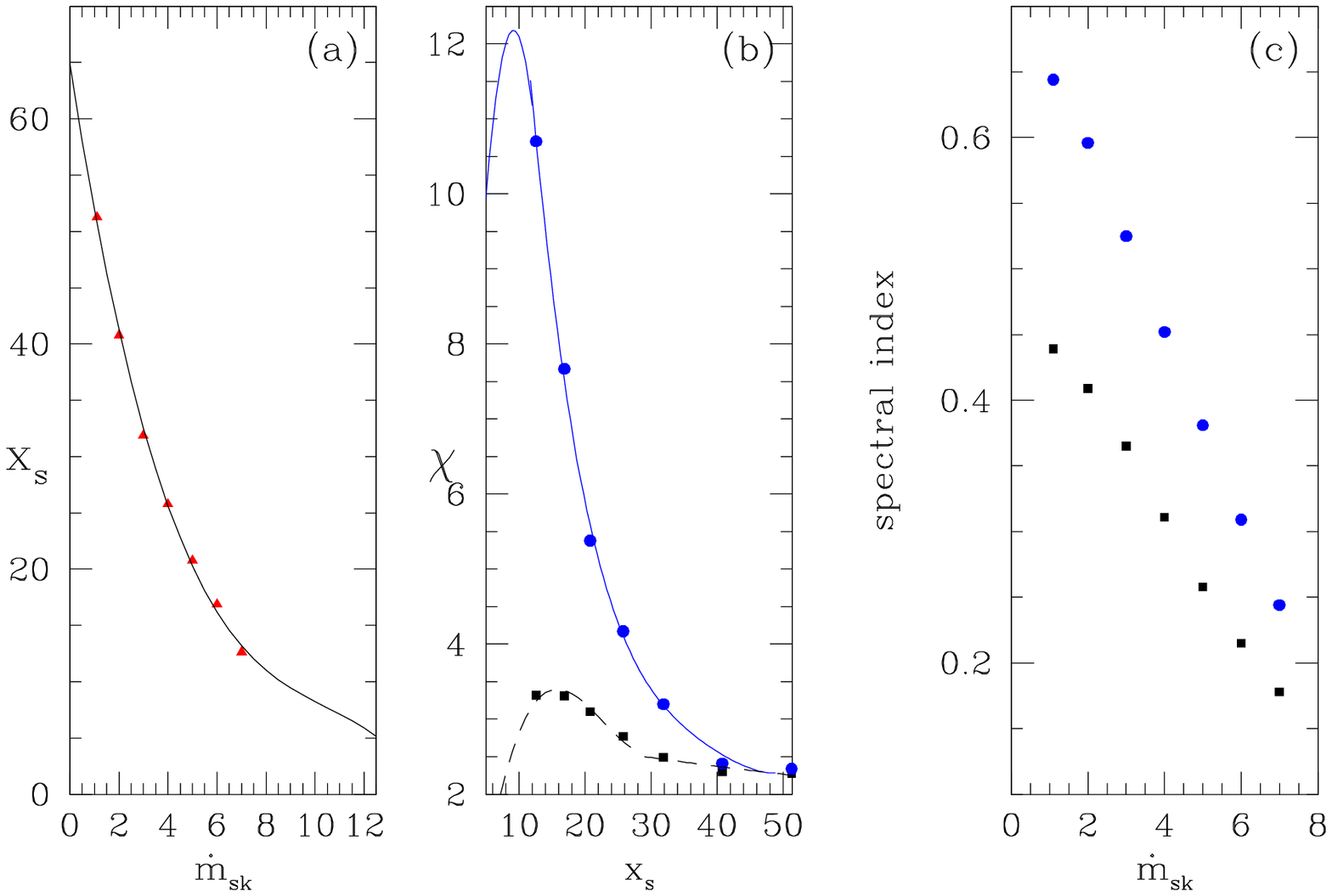}
 \caption{(a) Shock location $x_s$ as a function of ${\dot m}_\sk$ (filled triangles, red online).
The fitted function given by equation (\ref{xsdotm.eq}) (solid curve, black online). Parameters
fixed to get the data (filled triangles) are $r_0=3686$, $\vartheta_0=1.928\times 10^{-3}$,
$\Theta_{0}=9.811\times 10^{-2}$,
$\lambda_0=1.7$ and $\alpha=0.001$. (b) $\chi$ as a function
of $x_s$ for $10 M_{\odot}$ (filled square, black online) and $10^8 M_{\odot}$ BH (filled circle, blue online).
The corresponding fitted functions for $10 M_{\odot}$ BH (dashed, black online) and $10^8 M_{\odot}$ BH
(solid, blue online) presented in equations (\ref{chi8.eq}, \ref{chi1.eq}).
(c) Spectral index as a function of ${\dot m}_\sk$ plotted for the same data points of Fig. a. for $10 M_{\odot}$ BH
(filled square, black online) and $10^8M_\odot$ BH (filled circle, blue online).}
 \end{center}
\label{lab:FigB1}
\end{figure}
In Fig. (\ref{lab:FigB1}1a) red triangles represent the locations of shock for a given $\dot m_\sk$.
Other parameters are $r_0=3686$, $\vartheta_0=1.928\times 10^{-3}$, $\Theta_{0}=9.811\times 10^{-2}$,
$\lambda_0=1.7$ and the viscosity parameter $\alpha=0.001$. In the figure, $x_s$ is computed using the methods of
\citet{kc14} for the given values of $\alpha$, $\dot m_\sk$, $r_0$, $\vartheta_0$, $\Theta_0$ and $\lambda_0$.
We fitted a curve using these data generated by \citet{kc14} and expressed $x_s$ as a function $\dot m_\sk$,
the explicit form of which is
\begin{equation}
x_s=64.8735-14.1476\dot m_\sk+1.24286\dot m_\sk^2-0.039467\dot m_\sk^3.
\label{xsdotm.eq}
\end{equation}
Therefore, while calculating the radiative moments, $x_s$ is no more a free parameter, but is estimated 
using equation (\ref{xsdotm.eq}).

\subsection{Obtaining relation between shock location and ratio 
of pre-shock and post-shock luminosities}
\label{app:mdotxs2}

In previous works in which radiatively driven jets were studied \citep{cc00b,cc02,cdc04,c05},
the luminosity from PSD ($\ell_\s$) was supplied as a free parameter,
but in the present work, we calculate
this from spectral modeling.
KD produces the well known thermal radiation \citep{ss73}, while SKD emits
via bremsstrahlung and synchrotron emission. 
For the same set of outer boundary conditions, {\ie} $\alpha$, $\dot m_\sk$, $r_0$, $\vartheta_0$, $\Theta_0$ and
$\lambda_0$
and ${\dot m}_\kd$ at $r_0$, we solve the general radiative transfer equations similar to those by \citet{ct95,mc08}.
For the disc parameters of Fig. \ref{lab:FigB1}1a,
the shock location varied between $51.25 \geq x_s \geq 12.6$
as one varied $1.1 \leq \dot m_\sk \leq 7$, keeping all other boundary conditions same.
The flow solutions of \citet{kc14} were obtained in a unit system where $M_B=1$, so the actual
luminosity and spectra will depend on the mass of the central object.
PSD
being optically slim will inverse-Comptonize radiations coming from pre-shock disc (SKD and KD).
Let the luminosity
from SKD and KD be denoted as $\ell_\s$ and $\ell_\kd$. Then the ratio of luminosities is defined as
$\chi=\ell_\s/(\ell_\sk+\ell_\kd)$. In Fig. (\ref{lab:FigB1}1b) we plot $\chi$
with $x_s$ for the same set of data points as in Fig. (\ref{lab:FigB1}1a) and considering ${\dot m}_\kd=1$, we obtain the
spectra and luminosities of each disc component for
$M_B=10M_{\odot}$ (black, squares) and $M_B=10^8M_{\odot}$ (blue, circles).

The behaviour of $\chi$ for $10^8 M_\odot$ is different from that of $10 M_\odot$ because synchrotron 
cooling is very efficient for a stellar mass black hole. For a $10 M_\odot$ black hole, the post-shock
luminosity increases initially as the post-shock flow is still very hot but as shock location moves close to black hole, 
$\dot{m}_\sk$ is very high and pre-shock synchrotron cooling becomes large enough to reduce the ratio ($\chi$).
On the other hand $\chi$ keeps on increasing for $10^8 M_\odot$ because Comptonization due to Keplerian soft photons is the
most efficient cooling process and as shock moves in, the supply of hot electron increases (as $\dot{m}_\sk$ increases)
which enhances the post-shock luminosity. 
In Fig. (\ref{lab:FigB1}1c) we plot the variation of spectral index 
with $\dot m_\sk$. We see that 
as $\dot{m}_\sk$ increases the spectral states becomes harder because supply of hot the electron increases. Moreover,
the spectral
index for a super-massive black hole is generally softer than a stellar mass black hole as the PSD of stellar mass
black hole is relatively hotter. We fitted the plots of Fig. (\ref{lab:FigB1}1b) with analytic functions
(the constants are written correct up to three decimal points) given by
\begin{eqnarray}
\chi_{8}=25.944-1.667 x_s+3.992\times 10^{-2} x_s^2-3.199\times 10^{-4} x_s^3~(x_s > 12); \\ \nonumber
\chi_{8}=1.449+2.336 x_s-0.127 x_s^2. ~(x_s \leq 12)
\label{chi8.eq}
\end{eqnarray}
\begin{eqnarray}
\chi_{1}=-2.525+0.913 x_s -4.438 \times10^{-2} x_s^2+6.522\times 10^{-4} x_s^3 ~(x_s\leq 29);
\\ \nonumber
\chi_{1}=-2.914-1.622\times 10^{-2} x_s +7.265\times 10^{-5} x_s^2+1.278\times 10^{-7} x_s^3 ~(x_s>29).
\label{chi1.eq}
\end{eqnarray}
Here, $\chi_8$ signifies the ratio of PSD to SKD+KD luminosity for $M_B=10^8M_{\odot}$ (blue, solid curve Fig.
\ref{lab:FigB1}1b),
while $\chi_1$ is the ratio of luminosities for $M_B=10M_{\odot}$ (black, dashed curve Fig. \ref{lab:FigB1}1b).
Therefore we supply ${\dot m}_\sk$ and parameters at the outer boundary and estimate the $\vartheta_\sk$, $\Theta_\sk$ and $\rho_\sk$.
up to the $x_s$ (obtained via equation \ref{xsdotm.eq}) and these solutions we obtain $\ell_\sk$. The $\ell_\kd$ can be estimated
from ${\dot m}_\kd$, $r_0$ (outer edge) and $x_s$ (inner edge). So from the preshock luminosity ($\ell_\sk+\ell_\kd$)
we can estimate the $\ell_\s$ using either expression (\ref{chi1.eq}) or (\ref{chi8.eq}) depending on the
central mass. In this work $10M_\odot$ is considered as a representative of stellar mass BH and $10^8M_\odot$
is considered as super massive BH.

\section{Equation of state}
\label{app:eos}
\begin {figure}
\begin{center}
 \includegraphics[width=10.cm]{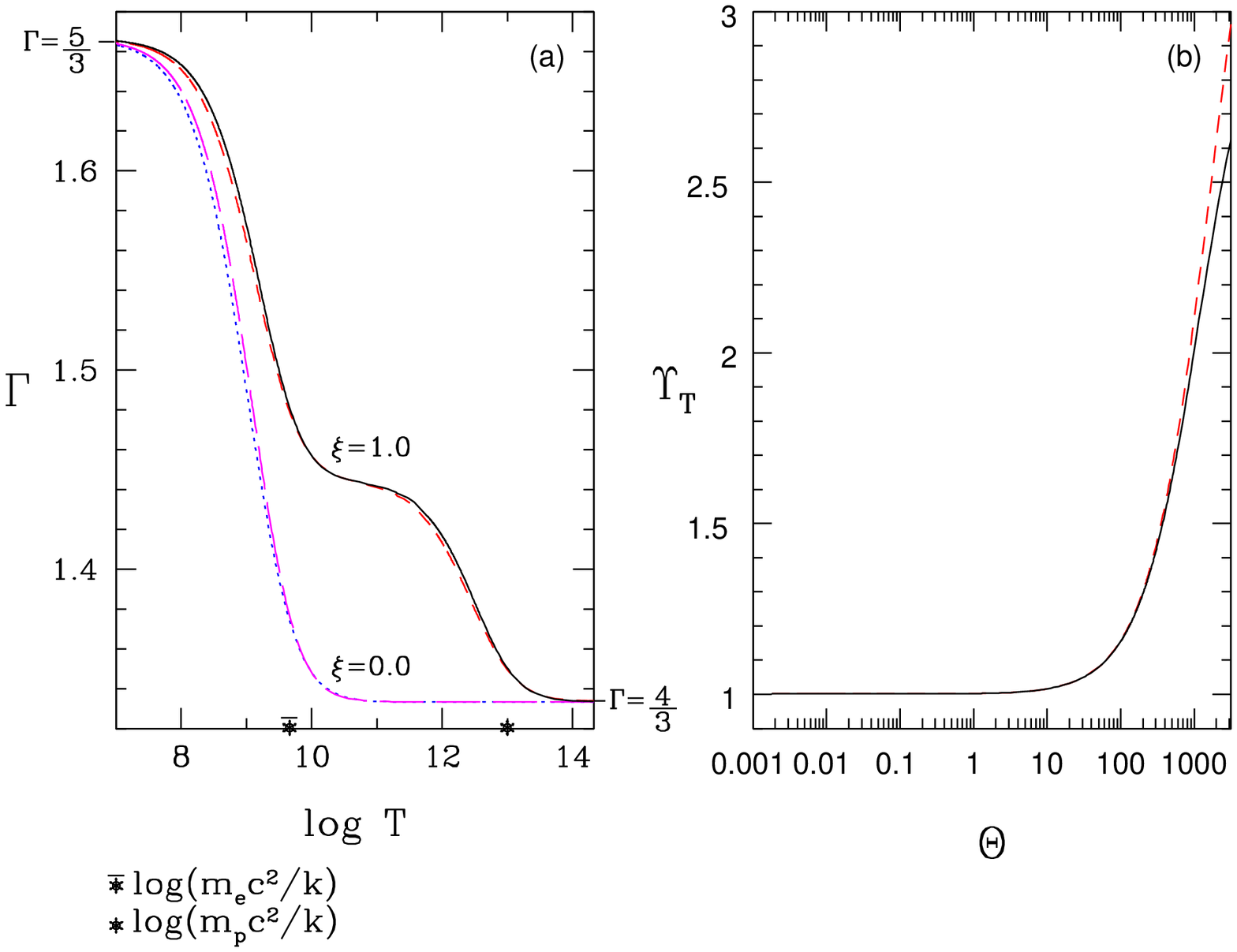}
\vskip -1.0cm
 \caption{(a) Comparing adiabatic index $\Gamma$ as a function of temperature $T$, for RP EoS (solid, black online and
long dashed, magenta online). CR EoS is presented by dashed curve (red online) and dotted (blue online) curves.
The two separate cases $\xi=0$ and $\xi=1$ are marked on the curves. Temperature corresponding to electron
and proton rest mass are also marked on the $T$ axis.
(b) $\Upsilon_T$ as a function of $\Theta$ for $\ep$ flow, but for RP EoS (solid, black online) and CR EoS
(dashed, red online).}
 \end{center}
\label{lab:FigC1}
\end{figure}
The EoS for relativistic fluid is obtained by integrating the relativistic energy of fluid particles
following a relativistic
Maxwell-Boltzman distribution in the momentum space as was obtained by \citet{c38, s57, cj68}. For single species fluid
the different forms of the EoS obtained by the above authors are as below,
\be
e_{\small \rm C}=\rho c^2\frac{3K_3(1/\Theta)+K_1(1/\Theta)}{4K_2(1/\Theta)};
~~e_{\small \rm S}=\rho c^2\frac{K_3(1/\Theta)}{K_2(1/\Theta)}-p;
~~ e_{\small \rm CG}=\rho c^2\left(3\Theta + \frac{K_1(1/\Theta)}{K_2(1/\Theta)}\right),
\label{rpeos1.eq}
\ee
where, $e$ represents local energy density of the flow, $\Theta$ the measure of temperature and suffix
C, S, and CJ signifies Chandrasekhar, Synge and Cox \& Giuli, respectively. The K's are modified Bessel's functions of second
kind and respective indices indicate their degree. We recall the recurrence relation $K_{m+1}(x)=K_{m-1}(x)+2m K_m/x$, and obtain
\be
h_{\small \rm C}=\frac{e+p}{\rho c^2}=\frac{3K_3}{4K_2}+\frac{K_1}{4K_2}+\Theta=\frac{K_1}{K_2}+4\Theta
=h_{\small \rm S}=h_{\small \rm CG}
\label{rpeos2.eq}
\ee
Therefore, all the forms of exact EoS for relativistic gas presented above, are equivalent, and let us denote
the EoS represented by equation (\ref{rpeos1.eq}) as `relativistically perfect' or RP EoS.
The multispecies approximate EoS used in this paper, may be called Chattopadhyay-Ryu or CR EoS (equation \ref{eos.eq}),
not only mimics equation (\ref{rpeos1.eq}) very well,
but also satisfies the fundamental inequality obtained by \citet{t48}. Taub showed from first principle that
any EoS for dilute relativistic gas have to satisfy a fundamental in-equality given by
\be
{\Upsilon}_{\rm T}=\left(h-\frac{p}{\rho c^2}\right)\left(h-\frac{4p}{\rho c^2}\right) \geq 1.
\label{taub.eq}
\ee
By following
equation (\ref{sound.eq}), we can calculate $\Gamma$ for any EoS.
In Fig. (\ref{lab:FigC1}1a) we plot adiabatic index $\Gamma$ as a function of temperature $T$,
for $\ep$ (marked $\xi=1$) and $\el$ (marked $\xi=0$) flow. For $\ep$ flow, the solid curve (black online)
is the $\Gamma$ with RP EoS, and dashed curve (red online) is the $\Gamma$ with CR EoS. And for $\el$
flow, $\Gamma$ with RP EoS is the long dashed (magenta online) curve, and that due to CR is dotted (red online) curve.
In Fig. (\ref{lab:FigC1}1b) we compare the Taub function $\Upsilon_T$ as a function of $\Theta$ for $\ep$ flow
described by RP EoS (solid, black online) and CR EoS (dashed, blue online). Both the EoS comfortably satisfy
the in-equality.

\section {On the jet geometry}
\label{app:jgeom}
\begin {figure}
\begin{center}
 \includegraphics[width=10.cm]{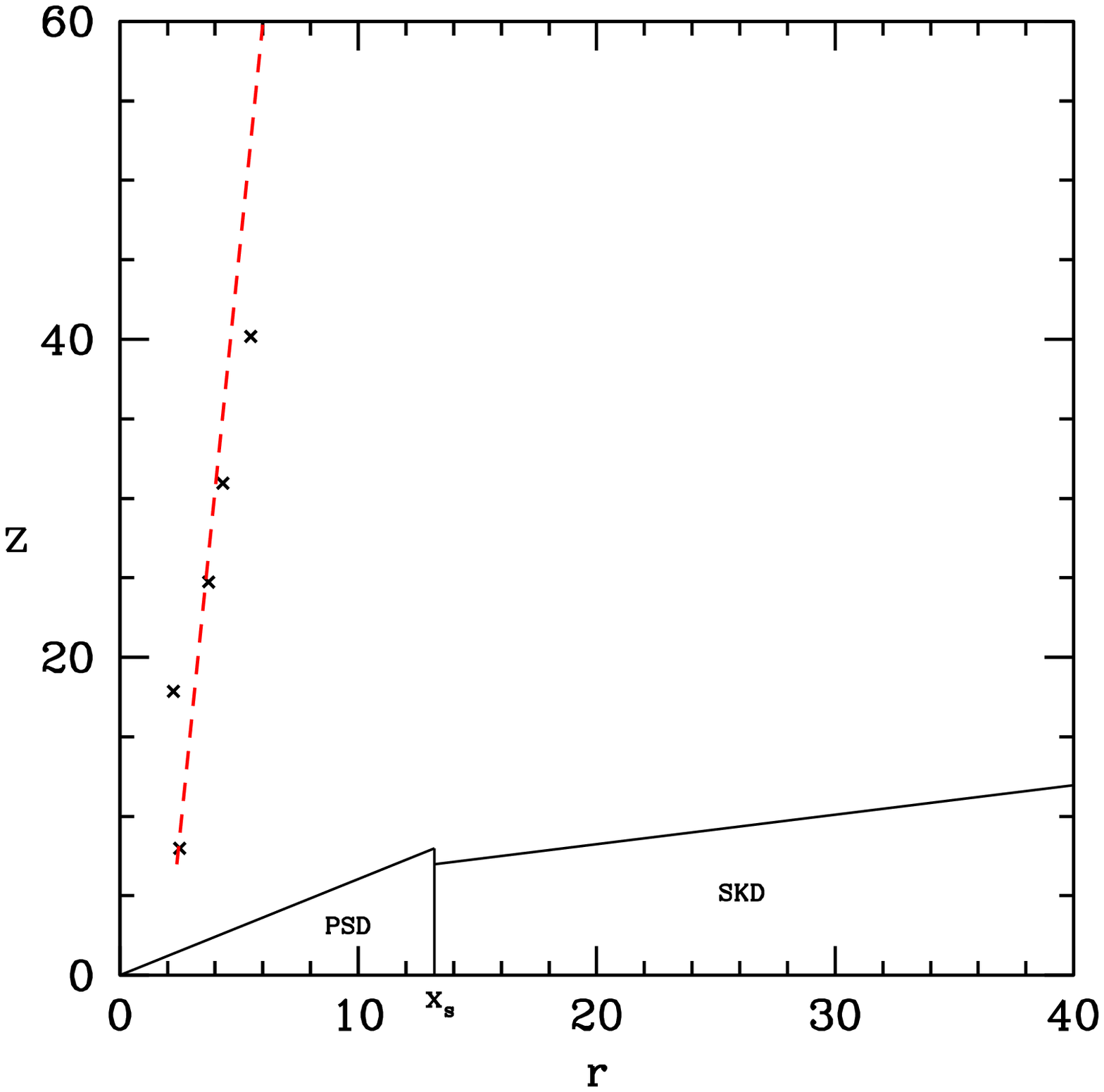}
 \caption{Estimated width of the jet by (crosses) by matching the thermal, gravity and radiative forces along $r$,
for jet base values $z_b=1.5,~a_b=0.50544,~v_b=0.00345$ and disk parameters ${\dot m}_{\sk}=7$ and ${\dot m}_\kd=1$.
The conical cross-section approximation is shown by a fitted line (dashed, red online).
The disc top surface and disc components (PSD and SKD) shown, is estimated from accretion rates supplied.}
 \end{center}
\label{lab:FigD1}
\end{figure}
In this paper we assumed the flow surface of the jet to be conical having a small opening angle and
maintains it throughout the spatial extent. Here we test how good is the approximation.
To approximately locate the lateral extent of the jet, we balance the pressure gradient term, the gravity term
and the radiative term along $r$ direction. We assume that $\partial p/(\rho \partial r)$ at the jet edge
is equal to $\partial p/(\rho \partial z)$ on the axis.
So
\begin{equation}
{\rm a}_{p_r}=-\frac{1}{\rho }\frac{\partial P}{\partial r}\approx \frac{2\Gamma \Theta}{\tau}\left(\frac{\gamma^2}{v}\frac{dv}{dz}+\frac{2}{z}\right)
\label{atr.eq}
\end{equation}
Similarly, approximating $u^r\approx u^{\phi} \approx 0$, the component of the radiative term along $r$
can be evaluated as (see, equation 3a of Chattopadhyay 2005),
\begin{equation}
{\rm a}_{r_r}=\gamma  \left({\cal F}^{r}-v {\cal P}^{rz}\right)
\label{arr.eq}
\end{equation}
The gravity term is
\begin{equation}
{\rm a}_{g_r}=\frac{r}{2R(R-1)^2};~~ R=(r^2+z^2)^{1/2}
\label{arr.eq}
\end{equation}
In Fig. (\ref{lab:FigD1}1) we plot the location where $|{\rm a}_{p_r}|=|{\rm a}_{r_r}+{\rm a}_{g_r}|$ for a radiation field
due to disc parameters $\dot m_{sk}=7$, $\dot m_{kd}=1$ and jet characterized by $z_b=1.5,~a_b=0.50544,~v_b=0.00345$.
The accretion disc upper surface is also shown in the figure.
The locations where $|{\rm a}_{p_r}|=|{\rm a}_{r_r}+{\rm a}_{g_r}|$, is shown by crosses. 
The approximated boundary of the jet (dashed, red online), shows consideration of
conical jet flow geometry is a fairly good assumption. Although, this assumption is reasonable
only when the jet opening angle is small. This is because for large opening angle,
jet material at the edges would be spun up by ${\cal F}^{\phi}$, which may contribute in spreading the jet.
\end{document}